\documentclass{aa}  

\usepackage{graphicx}

\usepackage{xcolor}
\usepackage{amssymb}
\usepackage{amsmath}
\usepackage{CJK}
\usepackage{ulem}
\usepackage{booktabs}
\usepackage{epstopdf}

\usepackage{txfonts}

\newcommand{\ssst}{\scriptscriptstyle}
\newcommand{\E}[1]{\times 10^{#1}}

\newcommand{\RA}[3]{{#1}^{{\rm h}}{#2}^{{\rm m}}{#3}^{{\rm s}}}
\newcommand{\decl}[3]{{#1}^{\circ}{#2}'{#3}''}
\newcommand{\RAdot}[4]{{#1}^{{\rm h}}{#2}^{{\rm m}}{#3}\fs{#4}}
\newcommand{\decldot}[4]{{#1}^{\circ}{#2}'{#3}\farcs{#4}}

\newcommand{\s}{\,{\rm s}}      \newcommand{\ps}{\,{\rm s}^{-1}}
    \newcommand{\Msun}{M_{\odot}}
\newcommand{\cm}{\,{\rm cm}}    \newcommand{\km}{\,{\rm km}}

\newcommand{\erg}{\,{\rm erg}}        
    
    \newcommand{\G}{\,{\rm G}}

\newcommand{\nel}{n_{e}}        \newcommand{\NH}{N_{\ssst\rm H}}

\newcommand{\Rs}{R_{\rm s}}

\newcommand{\nH}{n_{\ssst\rm H}}        \newcommand{\mH}{m_{\ssst\rm H}}

\newcommand{\dua}{d_{8.5}} 
\newcommand{\dub}{d_{3.1}} 
\newcommand{\duc}{d_{50}} 
    
\newcommand{\Rb}{R_{\rm b}}

\newcommand{\snra}{Kes~73}
\newcommand{\snrb}{RCW~103}
\newcommand{\snrc}{N49}

\begin{document}

   \title{Spatially resolved X-ray study of supernova remnants that host
   magnetars: Implication of their fossil field origin} 

   \author{Ping Zhou
          \inst{1,2}
          Jacco Vink \inst{1,3,4},
          Samar Safi-Harb \inst{5},
          and
          Marco Miceli \inst{6,7}
          }

\institute{Anton Pannekoek Institute, University of Amsterdam, PO Box 94249, 1090 GE Amsterdam, The Netherlands; \email{p.zhou@uva.nl}
\and School of Astronomy and Space Science, Nanjing University,
Nanjing~210023, China
\and GRAPPA, University of Amsterdam, PO Box 94249, 1090 GE Amsterdam, The Netherlands
\and SRON, Netherlands Institute for Space Research, Sorbonnelaan 2, 3584 CA Utrecht, The Netherlands
\and Department of Physics and Astronomy, University of Manitoba, Winnipeg, MB R3T 2N2, Canada
\and Dipartimento di Fisica e Chimica E.\ Segr\`{e}, Universit\`{a} degli Studi di Palermo, Palermo, Italy
\and INAF-Osservatorio Astronomico di Palermo, Palermo, Italy
}

\titlerunning{Spatially resolved X-ray study of supernova remnants that host magnetars}
\authorrunning{Zhou et al.}

\date{Received June 1, 2019; accepted July 15, 2019}

 
  \abstract
{
Magnetars are regarded as the most magnetized neutron stars in the Universe.
Aiming to unveil what kinds of stars and supernovae can 
create magnetars, we have performed a state-of-the-art 
spatially resolved spectroscopic X-ray study of the supernova 
remnants (SNRs) 
\snra, \snrb, and \snrc, which host
magnetars 1E~1841$-$045, 1E~161348$-$5055, and SGR~0526$-$66, respectively.
The three SNRs are O- and Ne-enhanced and are
evolving in the interstellar medium with densities of $>1$--$2~\cm^{-3}$.
The metal composition and dense environment
indicate that the progenitor stars are not very massive.
The progenitor masses of the three magnetars
are
constrained to be  $< 20~\Msun$ (11--$15~\Msun$ for \snra, 
$\lesssim 13~\Msun$ for \snrb,\ and 
$\sim 13$--$17~\Msun$ for \snrc). 
Our study suggests that magnetars are not
necessarily made from very massive stars,
but originate from stars that span a large 
mass range.
The explosion energies of the three SNRs 
range from $10^{50}~\erg$ to $\sim 2\E{51}~\erg$,
further refuting that the SNRs are energized by rapidly rotating (millisecond) pulsars.
We report that \snrb\  is produced by a weak supernova
explosion  with  significant fallback, as  
such an explosion explains the low explosion energy ($\sim 10^{50}~\erg$), small observed metal masses 
($M_{\rm O}\sim 4\E{-2}~\Msun$ and $M_{\rm Ne}\sim 6\E{-3}~\Msun$), and sub-solar abundances of 
heavier elements such as Si and S. 
Our study supports the fossil field origin
as an important channel to produce magnetars, given 
the  normal mass range ($M_{\rm ZAMS} < 20~\Msun$) of the progenitor stars, 
the low-to-normal explosion energy of the SNRs, 
and the fact that
the fraction of SNRs hosting magnetars is consistent 
with the magnetic OB stars with high fields.
}


   \maketitle
%

\section{Introduction}

Stars with mass $\gtrsim 8~\Msun$ end their 
lives with core-collapse (CC) supernova (SN)
explosions \citep[see][for a review]{smartt09}.
Two products are left after the explosion: a compact object (a neutron star, or a black 
hole for the very massive stars) and 
a supernova remnant (SNR). 
Both products are important sources relevant to numerous physical 
processes. Since the two objects share 
a common progenitor and are born in a 
single explosion, studying them together 
will result in a better mutual understanding 
of these objects and their origin.
  
Magnetars are regarded as a group of 
neutron stars with extremely high 
magnetic fields \citep[typically $10^{14}$--$10^{15}\G$, see][for a recent review and see references therein]{kaspi17}.
To date, around 30 magnetars and magnetar
candidates have been found in the Milky Way, Large Magellanic Cloud (LMC), and Small Magellanic Cloud \citep{olausen14}.
For historical reasons, these magnetars are categorised as anomalous X-ray pulsars and 
soft gamma-ray repeaters, based on their observational properties. However, the distinction between the 
two categories has blurred over the last 10--20 years.
Unlike the classical rotational powered
pulsars, this group of pulsars rotates
slowly with periods of $P\sim 2$--12~s, large
period derivatives $\dot{P}\sim10^{-13}$-- $10^{-10} \s\ps$, and are highly variable
sources usually detected in X-ray and 
soft $\gamma$-ray bands.
In recent years, the extremely slowly rotating
pulsar 1E 161348$-$5055 ($P=6.67$~hr)
in \snrb\ is also considered to be a magnetar,
because some of its X-ray characteristics (e.g., 
X-ray outburst) 
are typical of magnetars \citep{deluca06,li07,dai16,rea16,xu19}.

The origin of the high magnetic fields of magnetars is still an open question.
There are two popular hypotheses:
(1) a dynamo model 
involving rapid initial spinning 
of the neutron star
\citep{thompson93},
(2) a fossil field
model involving a progenitor star with
strong magnetic fields
\citep{ferrario06,vink06c,vink08c,hu09}.
The dynamo model predicts that
magnetars are born with rapidly rotating
proto-neutron stars (on the order of millisecond),
which can power energetic SN explosions
\citep[or release most of the energy through gravitational waves,][]{dallosso09}.
This group of 
neutron stars is expected to
be made from very massive stars 
\citep{heger05}.
The fossil field hypothesis predicts 
that magnetars inherit magnetic fields 
from stars with high magnetic fields.
Nevertheless, for the fossil field model,
there is still a 
dispute on whether
magnetars originate preferentially from  
high-mass progenitors \citep[$>20~\Msun$,][]{ferrario06,ferrario08} or 
less massive progenitors \citep{hu09}.

Motivated by the questions about the origin
of magnetars, we performed
a study of a few SNRs that host magnetars.
As the SNRs are born together with magnetars, studying them allows
us to learn
what progenitor stars and which kinds of
explosion can create this group of pulsars.
Therefore, we can use observations of SNRs to 
test the above two hypotheses. 

In order to get the best constraints of the 
progenitor masses, explosion energies,
and asymmetries of SNRs, we selected those 
SNRs showing  bright, extended X-ray emission.
Among the ten SNRs hosting magnetars \citep[nine in][and RCW~103]{olausen14},
only four SNRs fall into this category.
They are \snra, \snrb, \snrc\ (in the LMC), and
CTB~109.
CTB~37B is another SNR hosting a magnetar,
but with an X-ray flux one order of magnitude fainter and 
with sub or near-solar abundances \citep{yamauchi08,nakamura09,blumer19}.
Here we do not consider HB9, as the association 
between HB9 and the magnetar SGR~0501$+$4516 remains
uncertain.
\citet{vink06c} and \citet{martin14} have
studied the overall spectral properties
of SNRs \snra, \snrc, and CTB~109 and 
found that their SN explosions 
are not energetic.
In this study, with \snrb\ included and
CTB~109 excluded, 
we constrain the progenitor masses of 
the magnetars, provide spatial 
information about various parameters (such as abundances, temperature, density),
and explore the asymmetries
using a state-of-the-art binning method.
We exclude the oldest member CTB~109 
from our sample.\footnote{The X-ray emission in the western part of 
the SNR is almost totally absorbed, which
means that only a fraction of the metals 
can be observed.
For such an old SNR, the X-ray emission is 
highly influenced by the ISM.
The spectra are dominated by two thermal components, and therefore the derived metal
abundances and masses will be influenced  
by the assumed filling factors
of the X-ray-emitting gas.
Moreover, it might be difficult to constrain 
the age with good accuracy \citep[e.g., 9--14~kyr,][]{vink06c,sasaki13}. }
Therefore, our sample contains \snra, \snrb,
and \snrc, which host magnetars 
1E~1841$-$045, 1E~161348$-$5055, and SGR~0526$-$66, respectively.
Their ages have been well constrained, and 
the spectra of most regions could be 
well explained with a single thermal plasma model (see Sect.~\ref{sec:results}).
The distance of \snra\ is suggested to be
7.5--9.8~kpc using the HI observation by 
\citet{tian08} and 9~kpc using CO observation
\citep{liu17}. 
Here we take the distance of 8.5~kpc for \snra.
The distance of \snrb\ is taken to be 3.1~kpc
according to the HI observation 
\citep[the upper limit distance is 4.6~kpc]{reynoso04}. 
\snrc\ in the LMC is at a distance of 50~kpc.

\section{Data and method}

\subsection{Data}
We retrieved Chandra data of three SNRs --- \snra, \snrb, and \snrc.
Only observations with exposure longer than 15~ks are used.
The observational information is tabulated in Table~1.
The total exposures of the three SNRs are 152 ks, 107 ks, and 114 ks,
respectively.

We used CIAO software (vers. 4.9 and CALDB vers. 4.7.7)\footnote{http://cxc.harvard.edu/ciao} to 
reduce the data and extract spectrum. 
Xspec (vers. 12.9.0u)\footnote{https://heasarc.gsfc.nasa.gov/xanadu/xspec}
was used for spectral analysis.
We also used DS9\footnote{http://ds9.si.edu/site/Home.html} and IDL (vers.\ 8.6) to visualize and
analyze the data.

\begin{table}
\footnotesize
\caption{Observational information of SNRs that host magnetars.}
\label{T:info}
\begin{tabular}{lcccc}
  \hline\hline
SNRs & obs.\ ID & exposure (ks)  & obs.\ time & PI \\
\hline
Kes~73  & 729    & 29.6 & 2000-07-23 & Slane \\
        & 6732   & 25.2 & 2006-07-30 & Chatterjee\\
        & 16950  & 29.0 & 2015-06-04 & Borkowski\\
        & 17668  & 21.2 & 2015-07-07 & Borkowski\\
        & 17692  & 23.6 & 2015-07-08 & Borkowski\\
        & 17693  & 23.1 & 2015-07-09 & Borkowski\\
\hline
RCW~103 & 11823  & 63.3 & 2010-06-01 & Garmire  \\
        & 12224  & 18.1 & 2010-06-27 & Garmire  \\
        & 17460  & 25.1 & 2015-01-13 & Garmire  \\
\hline
N49     & 10123  & 28.2 & 2009-07-18 & Park \\
        & 10806  & 27.9 & 2009-09-19 & Park \\
        & 10807  & 27.3 & 2009-09-16 & Park \\
        & 10808  & 30.2 & 2009-07-31 & Park \\
\hline
\end{tabular}
\tablefoot{
For Kes~73 and RCW~103, the detector was ACIS-I. For N49, the
detector used was ACIS-S.
}
\end{table}

\subsection{Adaptive binning method}
In order to perform spatially resolved X-ray spectroscopy, 
we dissected the SNRs into many small regions and
extracted the spectrum from each region in individual observations. 
We employed a state-of-the-art adaptive spatial binning method called the weighted Voronoi tessellations  (WVT) binning algorithm \citep{diehl06},
a generalization of the \citet{cappellari03} Voronoi binning algorithm,
to optimize the data usage and spatial resolution.
The same method has been used to 
analyze the X-ray data of SNR W49B and study 
its progenitor star \citep{zhou18a}.
The X-ray events taken from the event file are adaptively binned to ensure that each bin 
contains a similar number of X-ray photons.
Therefore, the WVT algorithm allows us to
obtain spectra across the SNRs with
similar statistical qualities.

\begin{figure*}
  \centering{
  \includegraphics[angle=0, width=0.3\textwidth]{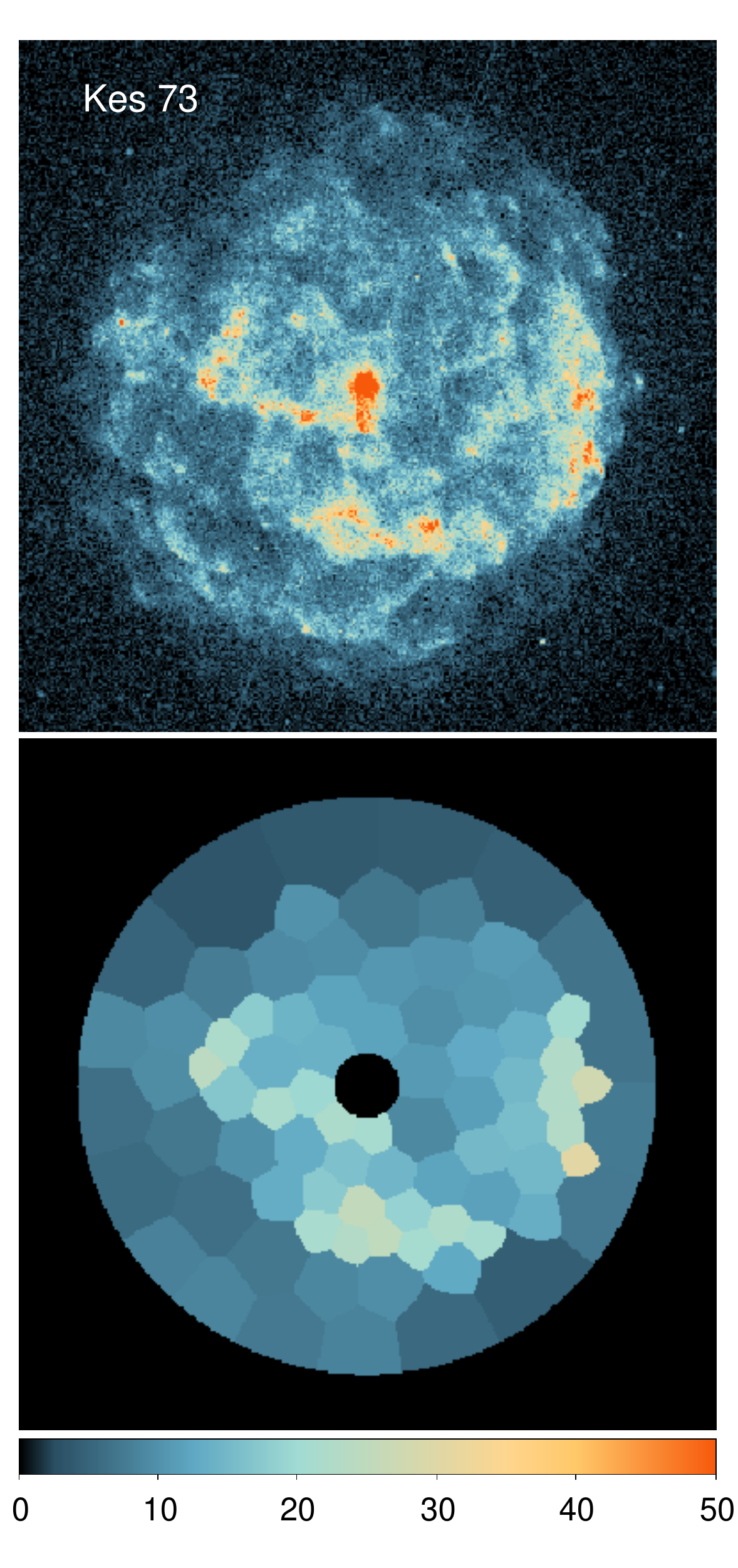}
  \includegraphics[angle=0, width=0.3\textwidth]{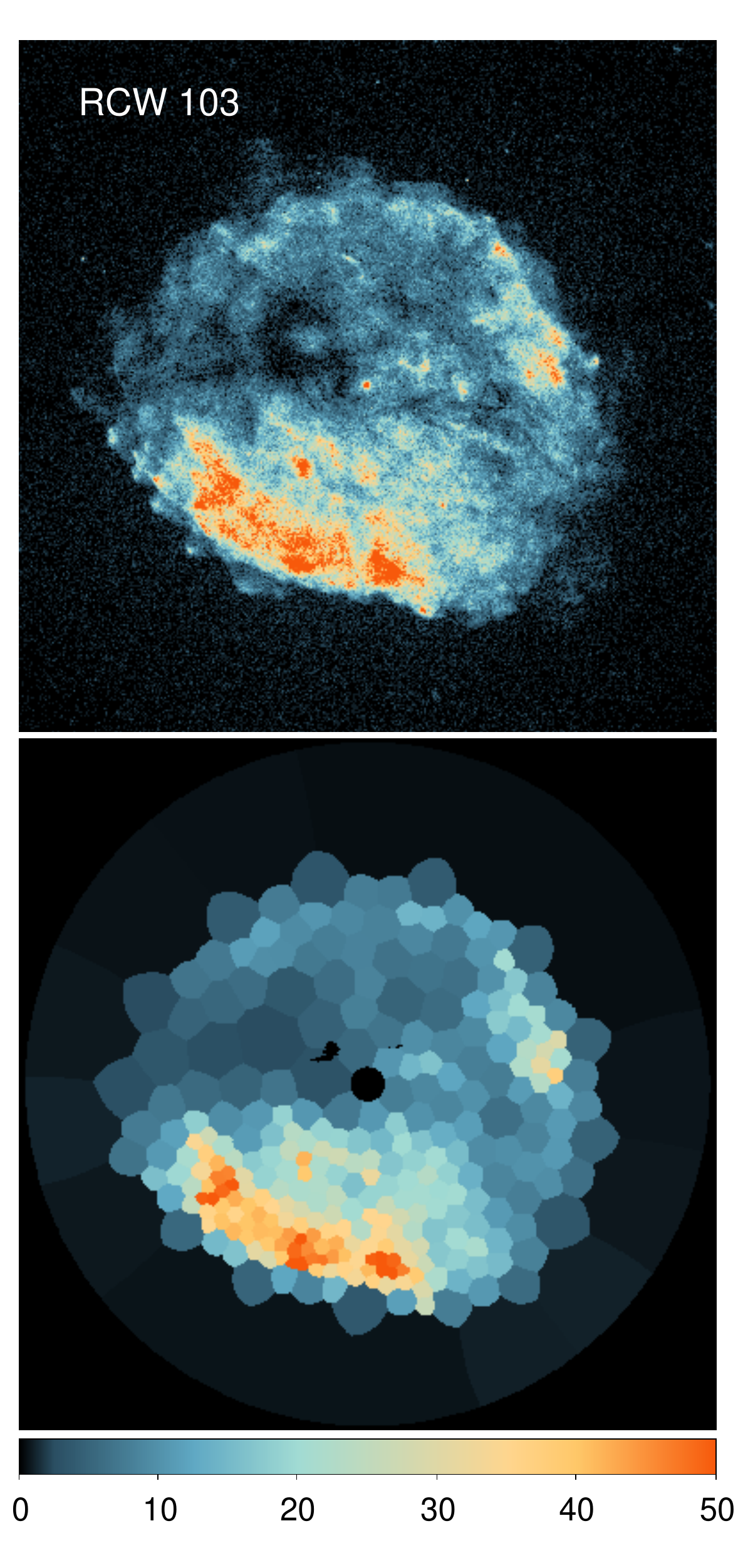}
  \includegraphics[angle=0, width=0.3\textwidth]{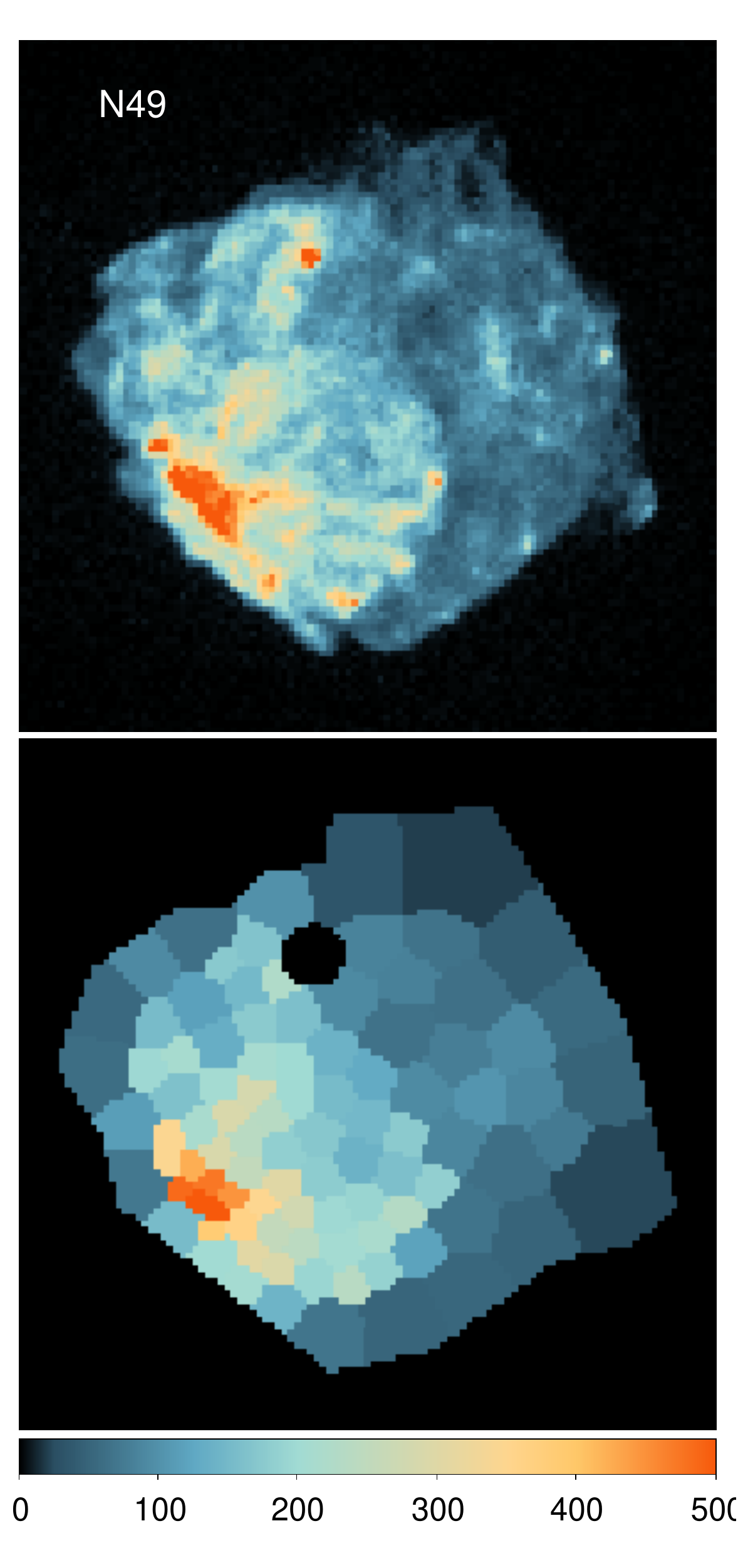}
  }
\caption{
Upper panels: the merged Chandra 
images of three SNRs in the 0.3--7.0 keV energy band. 
Lower panels: the adaptively binned
images with the magnetars removed. 
The colorbars
show the counts number per pixel ($1''$). 
}
\label{fig:wvtimg}
\end{figure*}

Firstly, for each SNR, we produce a merged 
0.3--7.0~keV image from all  observational epochs 
using the command merge\_obs in CIAO. 
This merged image is subsequently used to generate  spatial bins using the WVT algorithm.
Since this study focuses on the plasmas of SNRs,
we exclude the magnetars' emission by removing 
circular regions with angular radii of $15''$, 
$20''$, and $5''$ (radius to encircle over 95\% of the photon 
energy below 3.5~keV),  respectively, 
for \snra, \snrb, and \snrc.
We also exclude the pixels with an exposure short 
than 40\% of the total exposure. 
For the three SNRs, the targeted counts in each bin are 6400, 10000, and 
6400, respectively, corresponding to  signal-to-noise (S/N) ratios of 80, 100, and 80,  respectively. 
We obtain 83, 293, and 96 bins within 
\snra, \snrb, and \snrc, respectively.
Because \snrb\ is bright and is the most extended 
SNR among the three SNRs, we use a larger 
S/N to increase the statistics of each bin and 
do not define the SNR boundary. 
For the other two SNRs, we manually defined the boundaries of SNRs in 
order to include all the X-ray photons located
around the edges.
The merged images and adaptively binned images are shown in
Fig.~\ref{fig:wvtimg}.

Secondly, we extract spectra from each region (bin) in
individual observations and jointly fit the spectra
at each bin using a plasma model.
From spectral fit, we obtain the best-fit parameters and their uncertainties at different bins.
Finally, we study the distributions of the best-fit 
parameters across the SNRs and do further analysis
(see Sect.~\ref{sec:results}).

\section{Results} \label{sec:results}

\subsection{Spectral fit and density calculation}
\label{sec:fit}

The X-ray emission of the three SNRs can
be generally well fitted with an absorbed non-equilibrium ionization (NEI) 
plasma model, although 
some
regions might need double components
to improve the fit \citep[Miceli et al.\ in preparation, and][]{braun19}.
The plasma model uses the atomic data in the ATOMDB code \footnote{http://www.atomdb.org/}
version 3.0.9.
Using the single component model, we
consider that the SN ejecta and the ambient 
media
are mixed.
An appropriate NEI model to describe the shocked plasma 
in young  SNRs is the $vpshock$ model, which
describes an under-ionized plasma heated 
by a plane-parallel shock \citep{borkowski01}.
This model allows us to fit the electron temperature $kT$, metal abundances, and the ionization 
timescale $\tau=\int \nel t$, where
$\nel$ is the electron density and $t$ is the 
shock age (approximate  SNR age).
The Tuebingen-Boulder interstellar medium (ISM) absorption model $tbabs$ is used for calculation of the X-ray
absorption due to the gas-phase ISM, the grain-phase ISM, and the molecules in the ISM \citep{wilms00}. 
The solar abundances of \citet{asplund09} are adopted.
We note that both single-temperature
and multi-temperature components models are
frequently used in SNRs.
Two- or multi-temperature
components are often needed 
for large extraction regions characterized by mixed ejecta and blast wave components, which as a result show
a spatial variation of their spectral properties (such as the column density, the plasma temperature, the ionization timescale, or the gas density).
Here we performed a state-of-the-art spatially
resolved spectral analysis to address this complication.
If two-temperature components are indeed needed everywhere in the SNR, the final best-fit parameters might be affected.
Fitting a multi-thermal plasma with a single temperature causes a systematic 
error in the derived abundances. For example, an element whose strong lines have  emissivities that peak at the derived temperature may have its abundance underestimated, while an 
element whose lines peak away from the derived 
temperature will have an overestimated abundance. 
Although more complicated models are indeed needed in many SNRs, 
the spectral decomposition is generally nonunique
\citep{borkowski17} for many X-ray data and
uncertainties are difficult to account for.
The major reason for us to use the single
thermal component is that it gives an adequately
good fit to the spectra of most regions
\citep[in agreement with that pointed out by][for \snra]{borkowski17}.

Given the different spectral properties 
and environment of the three SNRs, 
the constrained metals are different.
When the abundance of an element cannot 
be constrained, we fix it to the value of
its environment (e.g., solar value in Kes~73 and
RCW~103; LMC value in N49).
For \snra, we fit the abundances of 
O (Ne tied to O), Mg, Si, S,  and Ar. 
The soft X-rays of \snrb\ and \snrc\ 
suffer less absorption,
allowing us to fit the abundances 
of O and Fe (Ni tied to Fe), in addition
to Ne, Mg, Si, and S.
\snrc\ is located in the LMC, so we used two absorption 
models to account for the Galactic and LMC
absorption: tbabs (Gal) $\times$ tbvarabs (LMC). 
The H column density of the Galaxy towards
N49 is fixed to $6\E{20}~\cm^{-2}$ \citep{park12} and the 
absorption in the LMC is varied.
The LMC abundances of C (0.45), N (0.13), O
(0.49), Ne (0.46), Mg (0.53), Si (0.87), S (0.41), 
Ar (0.62), and Fe (0.59) are taken from \citet[see references therein]{hanke10}. For other elements,
an averaged value of 0.5 is assumed.
The spectral fit results are summarized in the top part of
Table~\ref{tab:pars}.

\begin{table*}
\begin{center}
\footnotesize
\caption{Best-fit results and uncertainties.}
\label{tab:pars}
\begin{tabular}{lccc}
  \hline\hline
Parameters & \snra & \snrb & \snrc \\
\hline
\multicolumn{4}{c}{bin-averaged values (range) based on spectral fits} \\
\hline
$\chi^2_\nu$ 
&  1.11 (0.89 -- 1.49)
& 1.24 (0.88 --1.91)
& 1.09 (0.91 -- 1.42) \\
d.o.f
& 330 (268 -- 397)
& 220 (176 -- 276)
& 233 (170 -- 293)
\\
$\NH~(10^{22}~\cm^{-2})$ 
& 3.8 (3.3 -- 4.8) & 0.89 (0.45 -- 1.48) 
& 0.11 (< 0.38)\\
$kT$  
& 0.96 (0.72 -- 1.43)  & 0.63 (0.28 -- 1.14) 
& 0.68 (0.50 -- 0.92) \\
$[$O]
& 1.72 (0.14 -- 6.42) & 1.48 (0.37 -- 4.90) 
& 0.95 (0.26 -- 2.76)\\
$[$Ne]
& = [O]  & 1.37 (0.50 -- 2.38) & 1.19 (0.40 -- 3.66) \\
$[$Mg]
& 1.12 (0.61 -- 2.55) & 0.96 (0.65 -- 1.74)
& 0.47 (0.14 -- 1.37)\\
$[$Si]
& 1.14 (0.71 -- 2.07) & 0.74 (0.43 -- 1.78) 
& 0.58 (0.24 -- 1.58) \\
$[$S]
& 1.27 (0.73 -- 2.21) & 0.72 (0.10 -- 8.62)
& 1.59 (0.25 -- 3.37) \\
$[$Ar]
& 0.96 (0.12 --2.54) & 1 (fixed) &  1 (fixed)  \\
$[$Fe]=[Ni]
&  1 (fixed) & 1 (fixed)  & 0.27 (0.15 -- 0.45)\\
$\tau$ ($10^{11} \s~\cm^{-3}$)
& 2.23 (0.95--6.03) & 7.02 ($> 4.38$) 
& 28.0 ($>2.26$)\\
\hline
\multicolumn{4}{c}{bin-averaged density, total values, and 1-$\sigma$ uncertainties} \\
\hline
$\nH~(\cm^{-3})$ 
& $7.3^{+0.5}_{-0.4}$ 
& $5.9\pm 0.2$
& $6.6\pm0.3$ \\
$M_{\rm SNR}~(\Msun)$ 
& $46^{+3}_{-2}$
& $12.8\pm 0.4$
& $200_{-10}^{+14}$ \\
$t_{\rm sedov}$ (kyr)
& $\sim 2.4$ & $\sim 2.1$ & $\sim 4.9$
\\
$E_0$ (erg)
& $\sim 5.4\E{50}$
& $\sim 1.0\E{50}$
& $\sim 1.7\E{51}$
\\
$F_{\rm X}~(0.5$--7~keV; $10^{-11} \erg)$
& 2.5
& 17.4
& 2.3 
\\
\hline
\multicolumn{4}{c}{mass-averaged values and 1-$\sigma$ uncertainties } \\
\hline
[O]
& $1.54\pm0.2$ & $1.53\pm 0.11$ & $0.96\pm 0.12$ \\
$[$Ne]
& = [O] & $1.35\pm 0.08$ & $1.21 \pm 0.16$\\
$[$Mg]
& $1.1\pm 0.1$ & $0.96\pm 0.05$ & $0.47\pm 0.06$\\
$[$Si] 
& $1.1\pm 0.1$ & $0.74\pm 0.05$ & $0.57\pm 0.07$\\
$[$S] 
& $1.3\pm 0.1$ & $0.74^{+0.18}_{-0.09}$ & $1.55\pm 0.20$\\
$[$Ar] 
& $1.0\pm 0.1$ &  1 (fixed) &  0.62 (fixed) 
\\
$[$Fe] = [Ni]
&  1 (fixed) &  1 (fixed) & $0.27\pm 0.03$\\
$M_{\rm O}$ ($\Msun$) 
& $0.14^{+0.05}_{-0.04}$
& $3.9^{+0.7}_{-0.6}\E{-2}$
& $0.26\pm 0.05$\\
$M_{\rm Ne}$ ($\Msun$)
& $3.0^{+1.1}_{-0.8}\E{-2}$ 
& $5.8\pm 0.6\E{-3}$
& $8.7\pm1.5\E{-2}$\\
$M_{\rm Mg}$ ($\Msun$) 
& $1.9^{+1.6}_{-1.1}\E{-3}$
& $\dots$ & $\dots$\\
$M_{\rm Si}$ ($\Msun$) 
& $3.8^{+1.1}_{-0.8}\E{-3}$
& $\dots$ & $\dots$\\
$M_{\rm S}$ ($\Msun$) 
& $3.9^{+0.6}_{-0.5}\E{-3}$
& $\dots$
& $2.9\pm0.4\E{-2}$\\
$M_{\rm Ar}$ ($\Msun$) 
& $\dots$ & $\dots$ & $\dots$
\\
$M_{\rm Fe}$ ($\Msun$) 
& $\dots$ & $\dots$
& $\dots$ \\
\hline
\hline
\end{tabular}
\tablefoot{
The "$\dots$" sign 
indicates that the ejecta mass cannot be
calculated because the abundance is lower
than the solar or LMC value.
}
\end{center}
\end{table*}

The density is estimated based on an assumption
of the volume or geometry for the X-ray emitting plasma.
For a uniform density and a shock compression ratio
of four, mass conservation suggests that for shell-type SNR with a radius of $R$, the shell should
have a thickness of
 approximately $\Delta R=1/12R$: $4\pi R^2\Delta R (4\rho_0)=4\pi/3 R^3 \rho_0$,
 where $\rho_0$ is the ambient density.
The shell geometry is used to estimate the mean 
density $\nH$ for a given bin, combining
the normalization parameter in 
Xspec ($norm=10^{-14}/(4\pi d^2) \int n_e n_H dV$, where $d$ is the distance, $n_e$ and $n_H$ are the electron and H densities in the volume $V$; $n_e$=1.2$n_H$ for fully ionized plasma).
If the X-ray gas fills a larger fraction of the volume
across the SNR ($1/12<f<1$), the derived $\nH \propto f^{-1/2}$.
So the assumed geometry only affects the $\nH$ by a 
factor of up to 3.5.

The centers of \snra, \snrb, and \snrc\ are 
taken to be 
($\RAdot{18}{41}{19}{27}$, $\decldot{-04}{56}{12}{65}$),
($\RAdot{16}{17}{36}{12}$, $\decldot{-51}{02}{35}{75}$),
and ($\RAdot{05}{25}{59}{46}$,
$\decldot{-66}{04}{56}{73}$), respectively.
The radii are $2\farcm{22}$, $4\farcm{34}$, and $0\farcm{65}$, respectively.

\subsection{Distribution of parameters}

\begin{figure*}
  \centering
\includegraphics[angle=0, width=0.9\textwidth]{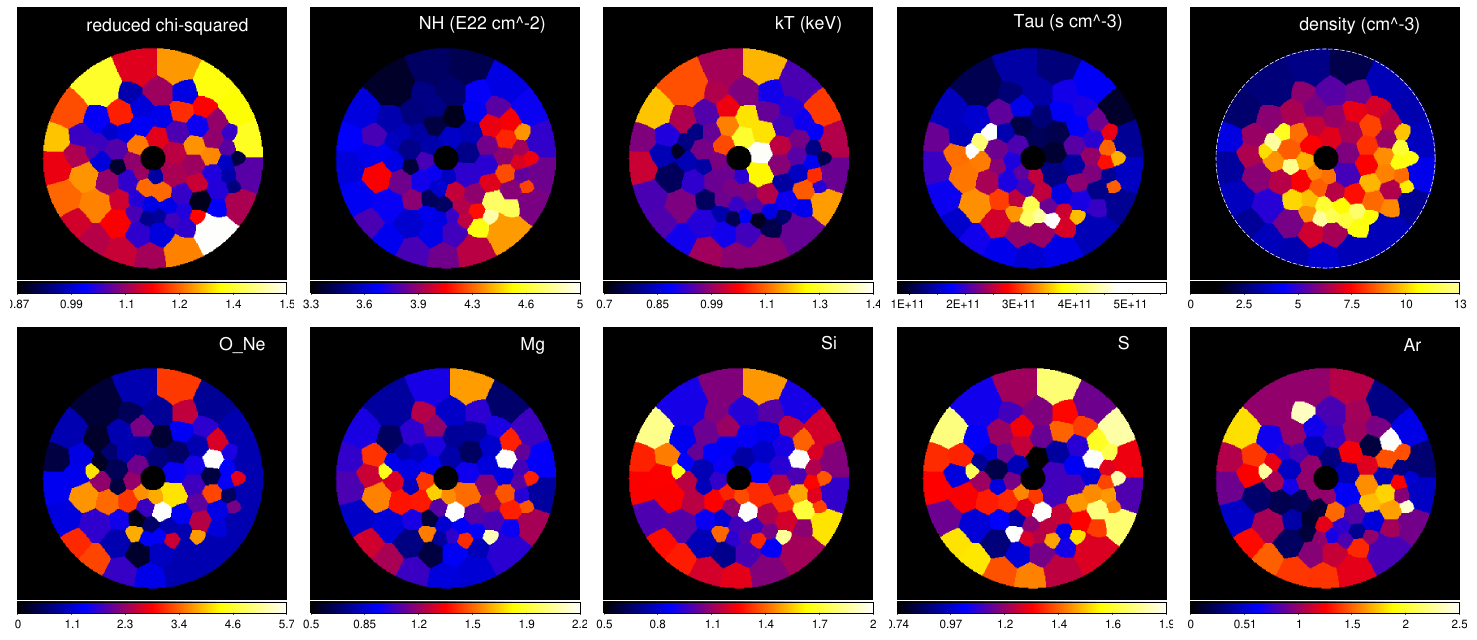}
\caption{
Distribution of the best-fit parameters in Kes 73.
The dashed circle indicates the outer boundary
for density calculation.
}
\label{fig:kes73_pars}

\vspace*{\floatsep}

  \centering
  \includegraphics[angle=0, width=0.9\textwidth]{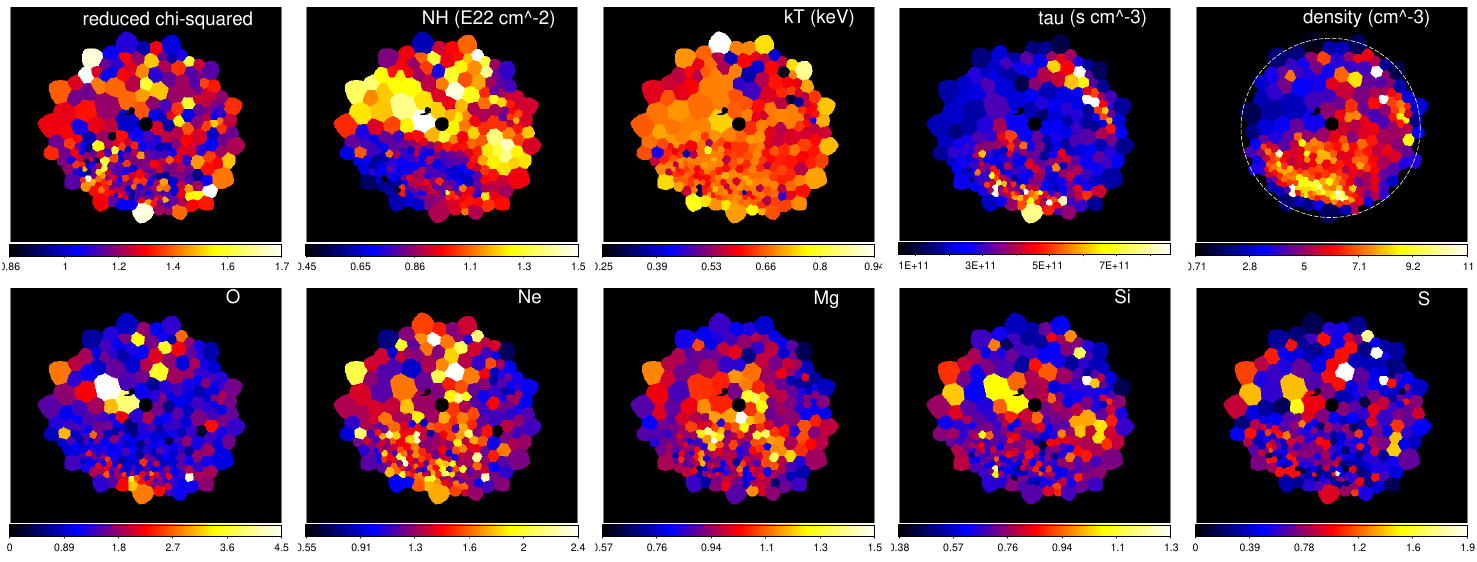}
\caption{
Distribution of the best-fit parameters in \snrb.
The dashed circle indicates the outer boundary
for density calculation.
}
\label{fig:rcw103_pars}

\vspace*{\floatsep}

  \centering
  \includegraphics[angle=0, width=0.9\textwidth]{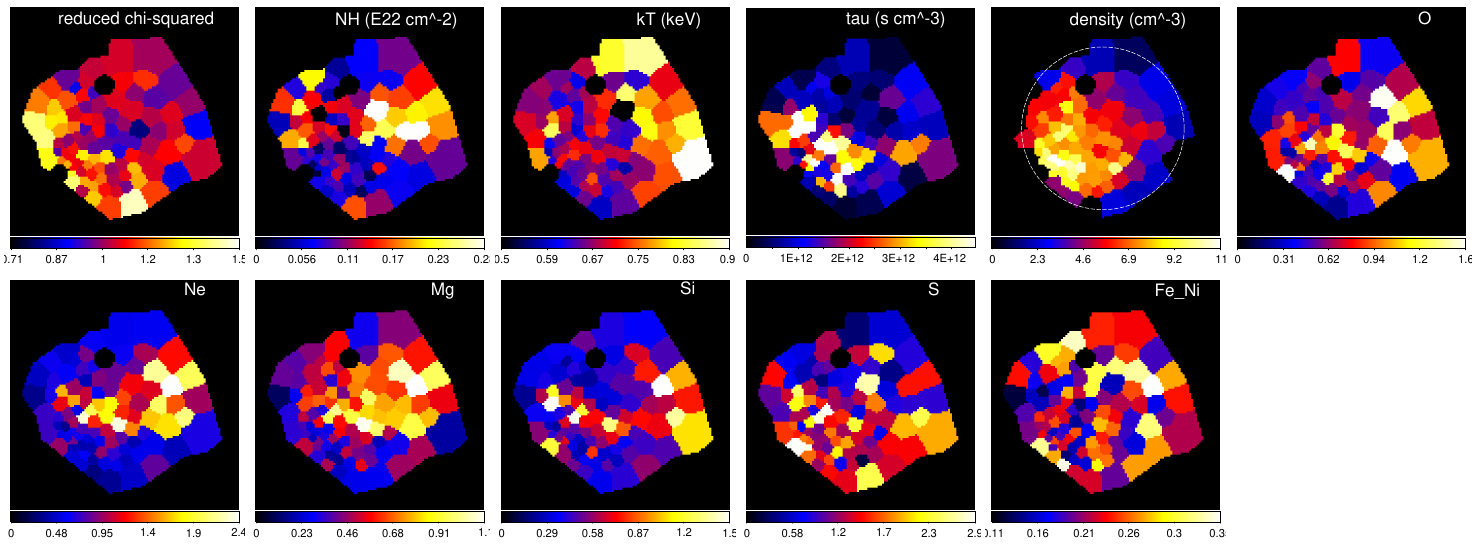}
\caption{
Distribution of the best-fit parameters in \snrc.
The dashed circle indicates the outer boundary
for density calculation.
}
\label{fig:n49_pars}
\end{figure*}

Figures~\ref{fig:kes73_pars}, \ref{fig:rcw103_pars}, and \ref{fig:n49_pars}
show the spatial distributions of the
best-fit parameters across the SNRs
\snra, \snrb, and \snrc, respectively,
except that the density panels are obtained
from the best-fit $norm$ and a geometry
assumption.
These three figures provide such ample 
information that it cannot be fully discussed 
in this work. 
In this paper, we will focus on the temperature,
metal content,  environment, and asymmetries.

\begin{figure*}
  \centering
  \includegraphics[angle=0, width=\textwidth]{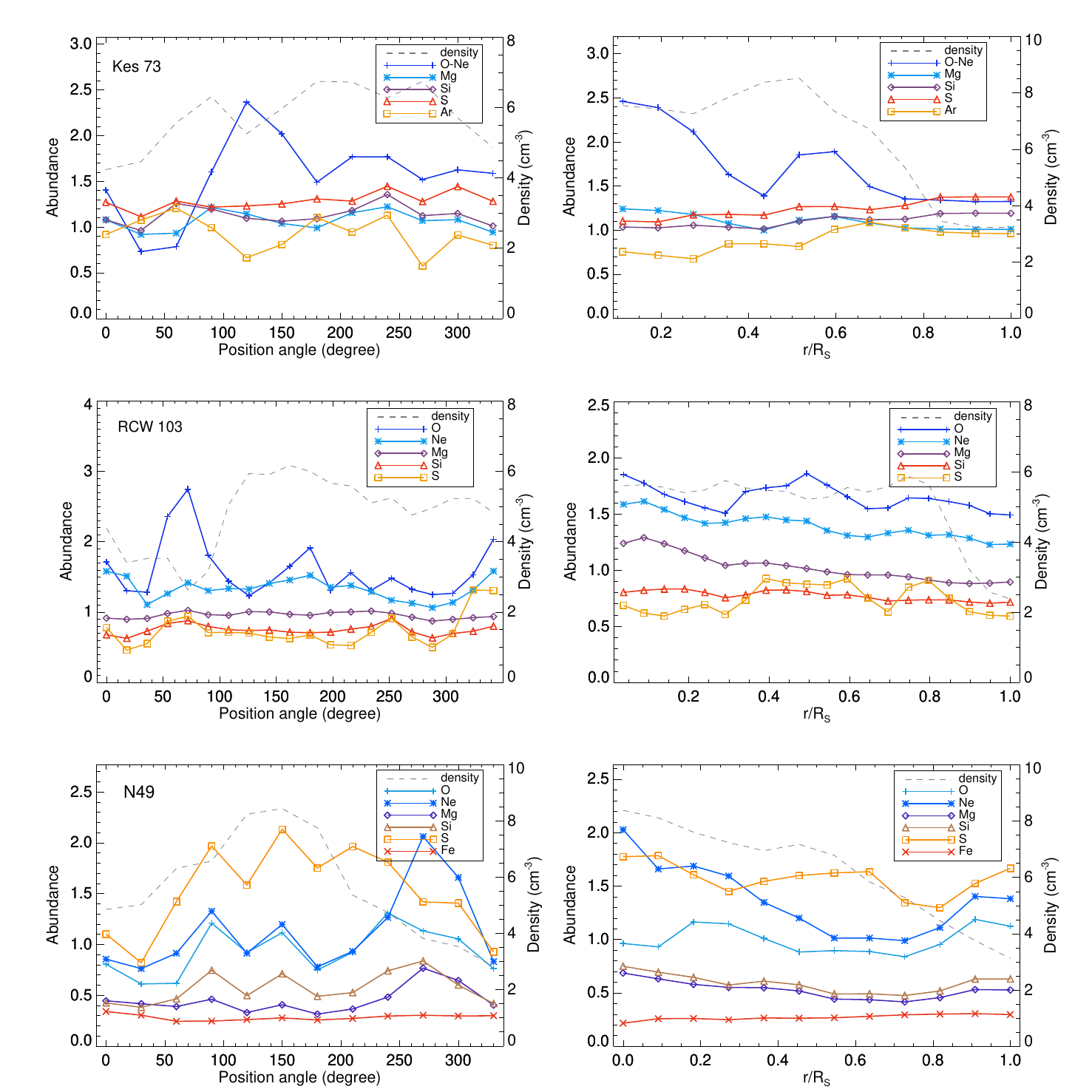}
\caption{
Azimuthal and radial profiles of the best-fit parameters in \snra\ (top),
\snrb\ (middle), and \snrc\ (bottom).
The position angle increases 
counterclockwise from the north where the
position angle is $0^\circ$. The SNR radius is
$R_{\rm S}$ .
The profiles of the post-shock density are 
plotted with dashed lines.
}
\label{fig:profile}
\end{figure*}

We also plot the azimuthal and radial distribution
of the best-fit parameters in Fig.~\ref{fig:profile}.
Here we briefly describe the distribution
of some important parameters.

\begin{itemize}
\item[--] \snra\ (Fig.~\ref{fig:kes73_pars}): 
There is a temperature variation across
the SNRs ($kT=$0.7--1.4~keV). 
The hottest plasma is located near the 
center of the SNR ($kT$ up to 1.4~keV), 
while there is a cold ($\sim 0.7$--0.8~keV),
broken-ring-like structure 
in the interior of the SNR
(ring radius of $~\sim 1\farcm{4}$, ring
centered at 
$\RA{18}{41}{18}$, $\decl{-04}{56}{13}$).
Such temperature variation is roughly
anti-correlated with the plasma density
and the X-ray brightness (see Fig.~\ref{fig:wvtimg}).

There are abundance enhancements of 
the O (Ne tied to O), Mg, Si, and S elements.
These elements show an east-west 
elongated structure, which is less 
clear in Ar, possibly because of the large
uncertainties of abundance [Ar] and that the
average [Ar] is less than one.
Another possibility is 
a result of the degeneracy
between [O] and $\NH$ in spectral
fit, as the higher [O] at some regions show slightly lower  $\NH$. 

Assuming that the gas is uniformly
distributed in each bin, the average
density is found to be $7.3^{+0.5}_{-0.4}~\cm^{-3}$,
suggesting an ambient density $n_0=\nH/4\sim 1.7~\cm^{-3}$,
consistent with the value obtained by \citet[$\sim 2\cm^{-3}$]{borkowski17}.
Such consistency indirectly supports that 
our geometry assumption is reliable to some extent.
The density is enhanced in a broken-ring-like structure
($\sim 10~\cm^{-3}$),
with an overall distribution similar
to that of the X-ray brightness.
\citet{liu17} suggested an interaction between
the SNR and a molecular structure in the east,
which may explain a larger column density $\NH$ there.

\item[--] \snrb\ (Fig.~\ref{fig:rcw103_pars}): The average temperature of
the X-ray-emitting plasma is $kT=0.63$~keV. 
The temperature distribution is nearly uniform,
except for a higher temperature in some boundary 
regions (outside the main shock sphere, 
likely related to high-speed ejecta clumps or
bad fit with single-temperature component model) and colder plasmas in the north.

We found that the O and Ne abundances are enhanced in \snrb,
while \citet{borkowski17} obtained near solar
abundances of them using the solar abundances of
\cite{grevesse98}, which give a 20\% lower solar
abundance of O compared to the abundance table
we used \citep[from][]{asplund09}.
The [Ne] and [Mg] are more enhanced in the SNR
interior, with an abundance gradient toward the
outer part (see Fig.~\ref{fig:profile}).
The element O is enhanced in the southern regions 
([O]$\sim 2$ in Fig.~\ref{fig:wvtimg}) 
and [O] is largest in two inner-east bins, 
with [O]=$5\pm1$ and $4\pm1$,  respectively. 
However, the $\NH$ values are
also highest at the two bins. 
Considering a degeneracy between $\NH$ and [O] 
when fitting the soft X-ray emission, the
[O] values at the two bins should be taken
with caution.

The average density is $5.9\pm 0.2~\cm^{-3}$.
The density distribution has a barrel shape.
The gas is greatly enhanced near the southeastern
boundary ($\nH\sim 9~\cm^{-3}$; $3\farcm{2}$ to the SNR center, $0\farcm{9}$ to the main shock boundary).

\item[--] \snrc\ (Fig.~\ref{fig:n49_pars}): There is an overall temperature gradient 
from the west to the east, anti-correlated with the 
density. The hottest bin is in the west,  with $kT=0.92$~keV.
The position is consistent with a protrusion as
shown in Fig.~\ref{fig:wvtimg}, which is a
Si and S-rich ([Si]=1.15 and [S]=2.3) ejecta knot.
A detailed study of the knot can be found
in \citet{park12}.

The O, Ne, and S elements are enriched in \snrc\ 
(mean [O]=0.96, [Ne]=1.2, and [S]=1.6) compared to the LMC values.
Similar to that in \snra\ and \snrb, the [Ne]
distribution is centrally enhanced.
Such an abundance gradient can be seen for
O and Mg as well.
Although the average abundances of Mg and Si are
smaller than LMC values, the two elements appear
more enhanced in the SNR interior. 
The distribution of S is asymmetrical, with higher
abundance in the south.

The average density is $6.6\pm 0.3~\cm^{-3}$.
There is a clear density gradient from the southeast
($\sim 10~\cm^{-3}$ ) to the northwest ($\sim 1~\cm^{-3}$).
This explains  why the X-ray emission is brightened
in the west.

\end{itemize}

\subsection{Global parameters} \label{sec:global}

Using the spectral fit results,
we calculate a few important parameters related
to the SNRs' evolution and metals:
gas mass $M_{\rm gas}$, metal mass 
$M_{\rm X}$,  SNR age $t$, and explosion energy $E_0$.
These results 
and the X-ray flux $F_{\rm X}$ in the 0.5--7~keV band are listed in Table~\ref{tab:pars}.

The masses of the X-ray-emitting gas $M_{\rm gas}$ are
calculated with the fitted $norm$ and assumed
geometry. Using a similar method for density
$\nH$, we derived total gas masses of $46^{+3}_{-2}~\Msun$
in \snra, 
$12.8\pm0.4~\Msun$ in \snrb, and $200^{+14}_{-10}~\Msun$ 
in \snrc.
We note that the assumed geometry of the density
distribution affects the $M_{\rm gas}$ by a factor
of a few.
If the X-ray gas fills a larger fraction 
of the volume across the SNR ($1/12<f<1$), 
the derived $M_{\rm gas}$ could be slightly
increased.
Therefore, if $f$ is assumed to be 1 (not likely for shell-type SNRs), we 
can derive the maximum limit of hot gas masses
of $61_{-3}^{+4}~\Msun$, $18.1^{+0.7}_{-0.5}~\Msun$,
and $260^{+17}_{-12}~\Msun$ for \snra, \snrb, and \snrc,
respectively.

The metal masses are important parameters that 
can be used to compare with the supernova 
yields predicted from nucleosynthesis models
and to test those models.
We obtain the mass-weighted average abundances
[X] and the observed ejecta mass $M_{X}$ 
as shown in the third part of Table~\ref{tab:pars}.
The abundance values are very similar to the bin-averaged 
abundances, so they are insensitive to the emission
volume assumptions.
The total masses of the metals are obtained by
summing up the metal masses in each bin. 
For element X, the mass is obtained as
$M_{X}=(([X]-[X]_{\rm ISM}) M_{\rm gas} f_{X}^m $,
where the interstellar abundance is $[X]_{\rm ISM}$=1 in
our Galaxy and is equal to the LMC value for \snrc,
and $f_{X}^m$ is the mass fraction of the element
in the gas.

The ages of the SNRs can be estimated from the electron
temperature $kT$ or from the ionization timescale $\tau$.
In the first method, the shock velocity is derived  as $v_{\rm s}=[16 kT_{\rm s}/(3
\mu m_{\rm H})]^{1/2}$, where
$\mH$ is the mass of hydrogen atom and $\mu=0.61$ is the mean atomic
weight for fully ionized plasma.
The relation between the shock velocity and 
the electron temperature holds in case of temperature equilibrium between different particle species 
(and this can be the case, considering the relative high values of $\tau$).
The Sedov age is $t_{\rm sedov}=2\Rs/(5v_{\rm s})$.
Using the averaged temperature $kT$ in these SNRs,
the ages of \snra, \snrb, and \snrc\ are found to be
2.4~kyr, 2.1~kyr, and 4.9~kyr, respectively.
These values are consistent with those obtained 
in previous expansion measurements
\citep[for \snrb\ and \snra, respectively]{carter97,borkowski17} and
X-ray studies \citep[for \snrc\ and \snra, respectively]{park12,kumar14}.
The X-ray emission of the three SNRs is characterized
by under-ionized plasma.
The shock age of an SNR $t$ can be inferred from the
ionization timescale $\tau=\int \nel t$ and the
gas density $\nH=1.2 \nel$ if the SNR is evolving in
a uniform medium.
We calculate a shock age in each bin and 
obtain the average age $\tau_{\rm shock}$ 
(range) of $0.9$~(0.4--1.8)~kyr  for \snra,
1.8 ($>0.6$) kyr for \snrb,  and 8.0 ($>0.8$)~kyr for \snrc. 
By comparing the $t_{\rm shock}$ values with $t_{\rm sedov}$
values, one would find that the difference is smallest
in \snrb, but much larger in \snra\ and \snrc, which are
evolving in very inhomogeneous environment (see density
distribution in Figs. \ref{fig:wvtimg} and \ref{fig:profile}). 
In a nonuniform medium, the $t_{\rm shock}$ may deviate from
the shock timescale.
Moreover, the $t_{\rm shock}$ values can be influenced by
the geometry assumption.
Therefore, we suggest that $t_{\rm sedov}$ better represents
the SNRs' true age $t$.

The explosion energy of an SNR in the Sedov phase
can be calculated using the Sedov-Taylor self-similar solution \citep{sedov59, taylor50, ostriker88}: 
$E_0=1/2.026\mu_{\rm H} \mH n_0 \Rs^5 t^{-2}$,
where $n_0$ is the ambient density.
We obtain $E_0\sim 5.4\E{50}\dua^{2.5}~\erg$,
$\sim 1.0\E{50}\dub^{2.5}~\erg$, and $\sim 1.7\E{51}\duc^{2.5}~\erg$ for \snra,
\snrb, and \snrc, respectively.
Here the radii are 5.5~pc, 3.9~pc, and 9.5~pc, respectively.

\section{Discussion}

The major goal of this paper is to explore which progenitor stars and which explosion mechanisms 
produce these SNRs and magnetars.
The explosion energy and metal masses 
are two important parameters to characterize the
explosion, while the distribution of metals 
provides information about the explosion (a)symmetries.
On the other hand, the density distribution provide
clues about  the environment and even 
mass-loss histories of the progenitor star.
In this section, we discuss these parameters
in order to unveil the explosion and progenitor
of magnetars.

\subsection{Environment and 
clues about the progenitor}
\label{sec:environment}

The density distribution is shown
in Figs.~\ref{fig:kes73_pars}, \ref{fig:rcw103_pars},
\ref{fig:n49_pars}, and \ref{fig:profile}, and gas 
masses are listed in
Table~\ref{tab:pars}.
The masses of the X-ray-emitting gas are $\sim 46~\Msun$ and
$\sim 200~\Msun$ in \snra\ and \snrc, respectively,
indicating that the gas is dominated by the 
ISM.
\snra\ is possibly interacting with molecular 
gas in the east \citep{liu17},
and \snrc\ is interacting with molecular 
clouds in the southeast \citep{banas97,otsuka10,yamane18}.
The inhomogeneous ambient medium significantly influences
the X-ray morphology of the SNRs.
As the density increases,  the X-ray emission 
is brightened.

Massive stars launch strong winds during
their main-sequence stage and could
clear out low-density cavities 
\citep[e.g.,][]{chevalier99}.
If the massive star is in a giant molecular cloud,
the maximum size of the molecular-shell bubble 
is linearly increased with the zero-age main-sequence stellar mass: 
$\Rb=1.22M_{\rm ZAMS}/M_\odot -9.16$~pc \citep{chen13}.
It is likely that the molecular shells 
found near the two SNRs were swept up by 
their progenitor winds.
Therefore, taking the distance of the 
molecular gas to the SNR center (about the
SNR radius), we can very roughly estimate 
the progenitor mass, which is $12\pm2~\Msun$
for \snra\ and $15\pm2$ for \snrc.
We note  here that the $\Rb$--$M_{\rm ZAMS}$ linear relationship
was obtained from the winds of Galactic 
massive stars, and may not be valid 
for LMC stars with lower metallicity.
Nevertheless, the derived mass of
\snrc\ agrees with the previous suggestion
of an early B-type progenitor ($M_{\rm ZAMS}<20\Msun$), which created
a 
Str\"{o}mgren 
sphere surrounding
\snrc\ \citep{schull85}.

The gas mass in \snrb\ is only $\sim 13~\Msun$.
Interestingly, the density distribution has a barrel shape with the shell at a distance of $\sim 3$~pc 
to the center (Fig.~\ref{fig:rcw103_pars}).
Between the position angle of $\sim 0^\circ$ (north) and $100^\circ$ (east), the density is 
reduced to 1/3--1/2 of that in other angles.
In the opposite direction, the density is
also relatively smaller.
The density is largest in the southern and northern
boundaries. 
This density distribution could explain why the
SNR is elongated (more freely expanding) toward the low-density direction.
It is likely that 
most of the X-ray-emitting gas has an ambient gas origin,
because  the density enhancement is consistent with
the distribution of molecular gas 
in the southeast, northwest, and west \citep{reach06}.
The existence of molecular gas with solar
metallicity \citep{oliva99} $\sim 3$--4~pc away
from the SNR center suggests that the progenitor
is not very massive ($M_{\rm ZAMS}< 13 \Msun$
if using the $\Rb$--$M_{\rm ZAMS}$ relationship).
Otherwise, the molecular gas would have been 
either dissociated by strong UV radiation or 
cleared out by fast main-sequence winds.

Although it is likely that the density
distribution reflects the ambient medium,
there is still a possibility that the lower density 
in the northeast and southwest is a
result of 
the pre-SN winds.
For example, a fast wind driven toward the 
northeast and  southwest may clear out 
two lower density lobes.
For a single star with $M_{\rm ZAMS}<15~\Msun$,
its red super-giant winds are generally slow ($\sim 10~\km\ps$) and 
the circumstellar bubble is small \citep[$<1$~pc][]{chevalier05}, which
cannot explain the low-density lobes.
If the progenitor star was in a binary system,
the accretion outflow could be fast and bipolar.
However, there is no observational evidence 
so far to support the progenitor being a
binary system.

\subsection{Explosion mechanism implied from the
observed metals} \label{sec:mechanism}

A common characteristic among the three SNRs is 
that all of them seem to be O- and Ne-enhanced, 
and there is no evidence of overabundant 
Fe (average abundance across the SNR).
\snrc\ reveals clear elevated [S], and \snra\ shows slightly 
elevated [S], but [S] is sub-solar in most
regions of \snrb.
Some regions with higher [O] show slightly lower $\NH$.
The degeneracy between the [O] and $\NH$ is difficult to distinguish using the 
current data.
Future X-ray telescopes with better spectroscopic 
capability and higher sensitivity may resolve the O lines and solve this problem.

\begin{figure*}
  \centering
{\includegraphics[angle=0, width=0.49\textwidth]{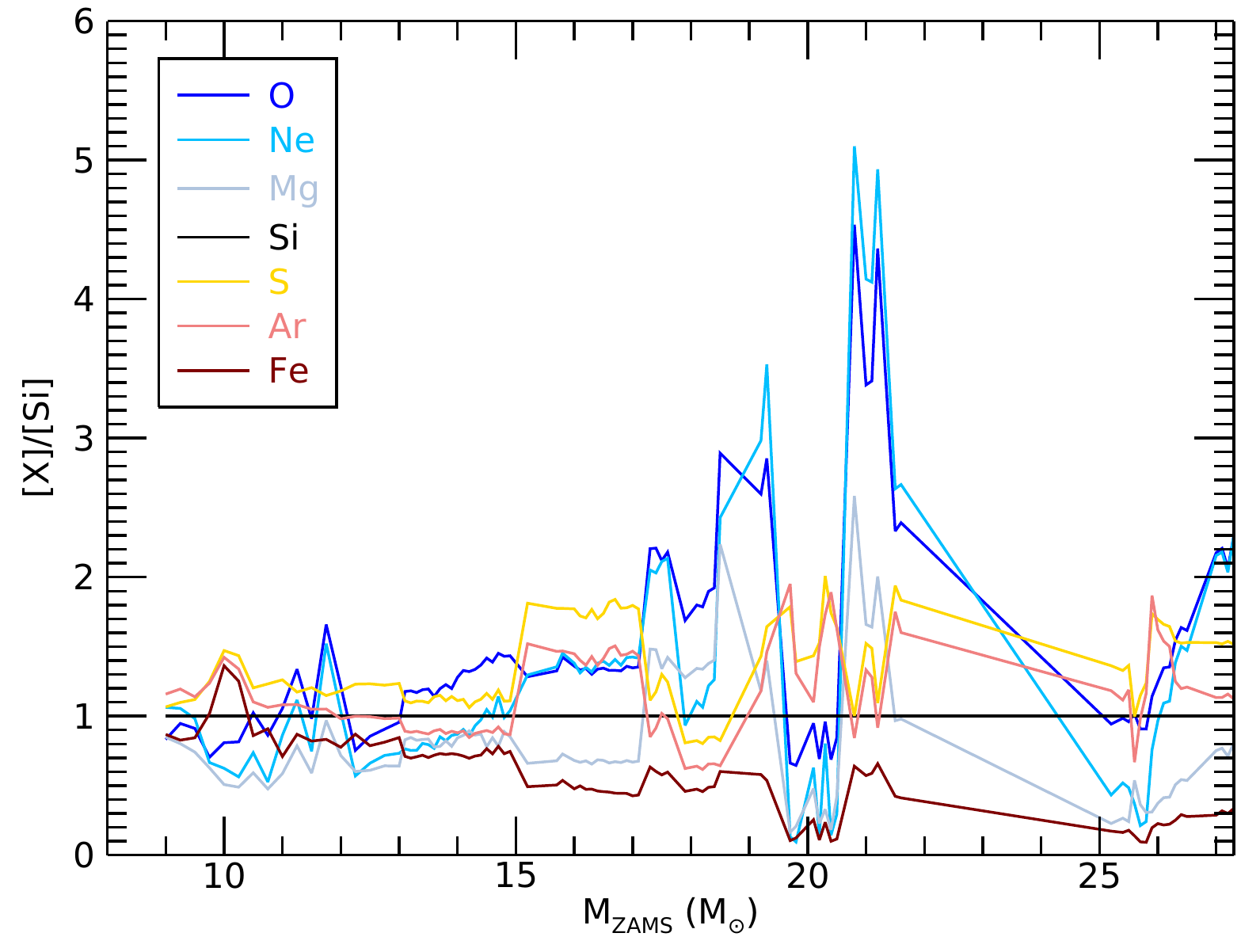}
\includegraphics[angle=0, width=0.49\textwidth]{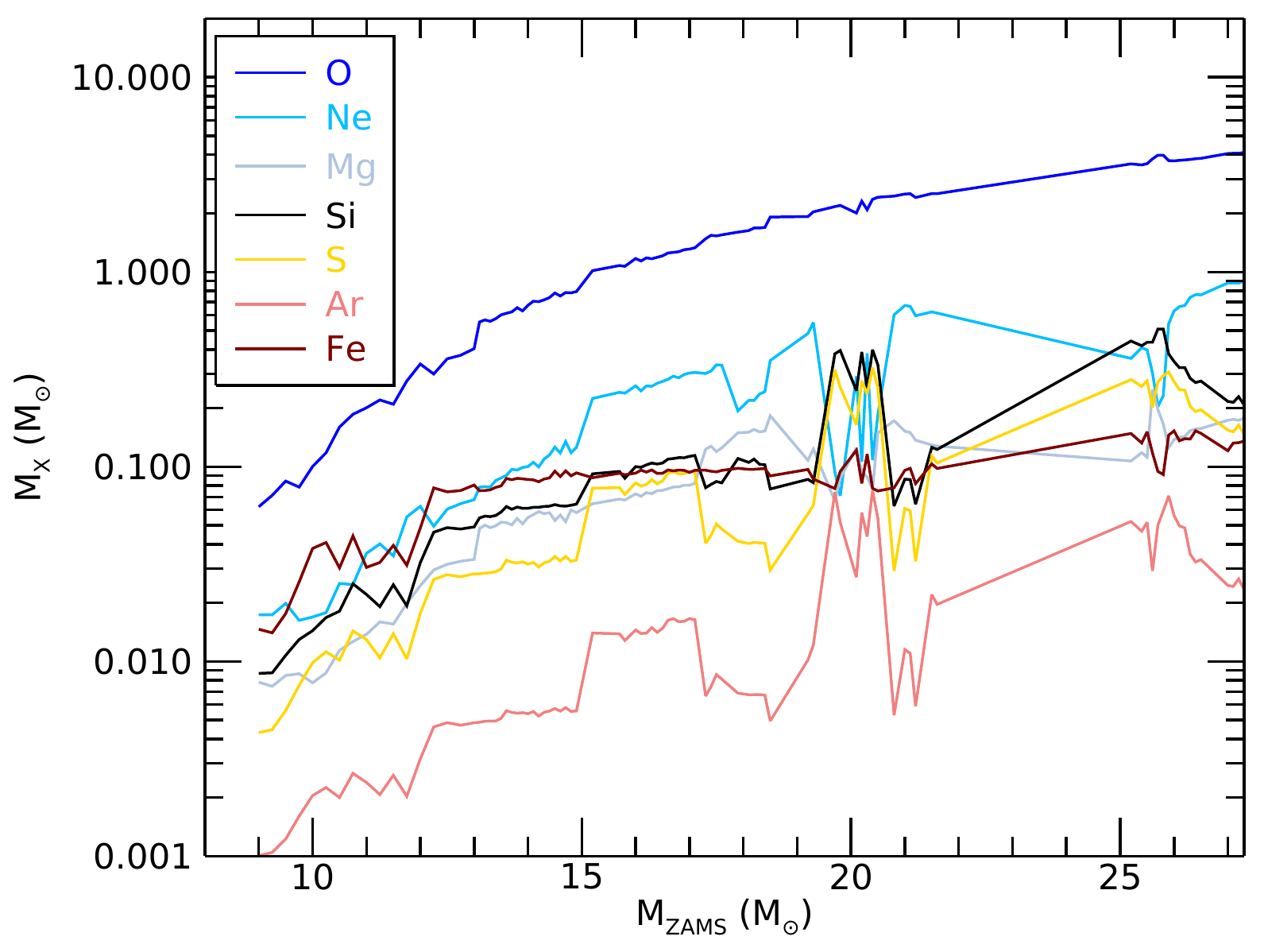}}
\caption{
Predicted abundance ratios (left) and yields (right) of the SN ejecta 
as a function of the progenitor masses based on the  
CC SN nucleosynthesis models by \citet{sukhbold16}.
The results of the 60~$\Msun$ and $120~\Msun$ stars are out of range of the plot. 
}
\label{fig:ratio}
\end{figure*}

\begin{figure*}
  \centering
{\includegraphics[angle=0, width=0.4\textwidth]{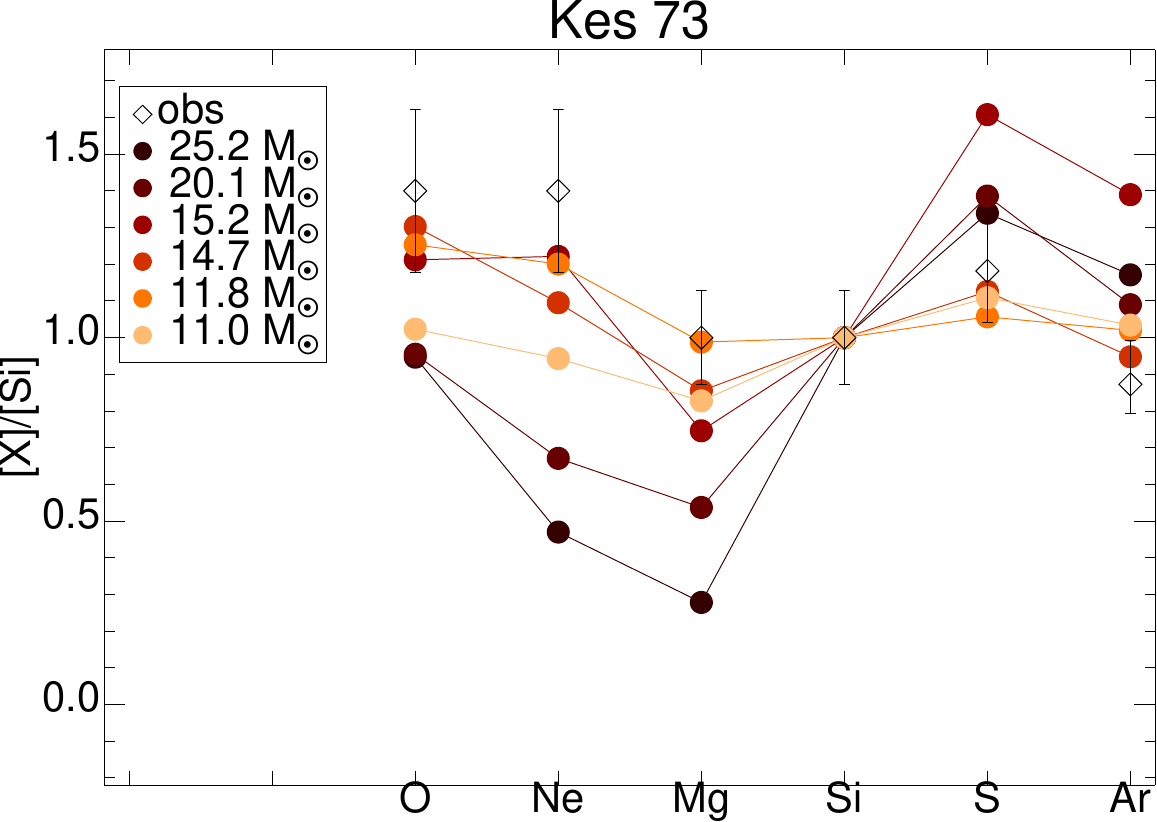}
\includegraphics[angle=0, width=0.4\textwidth]{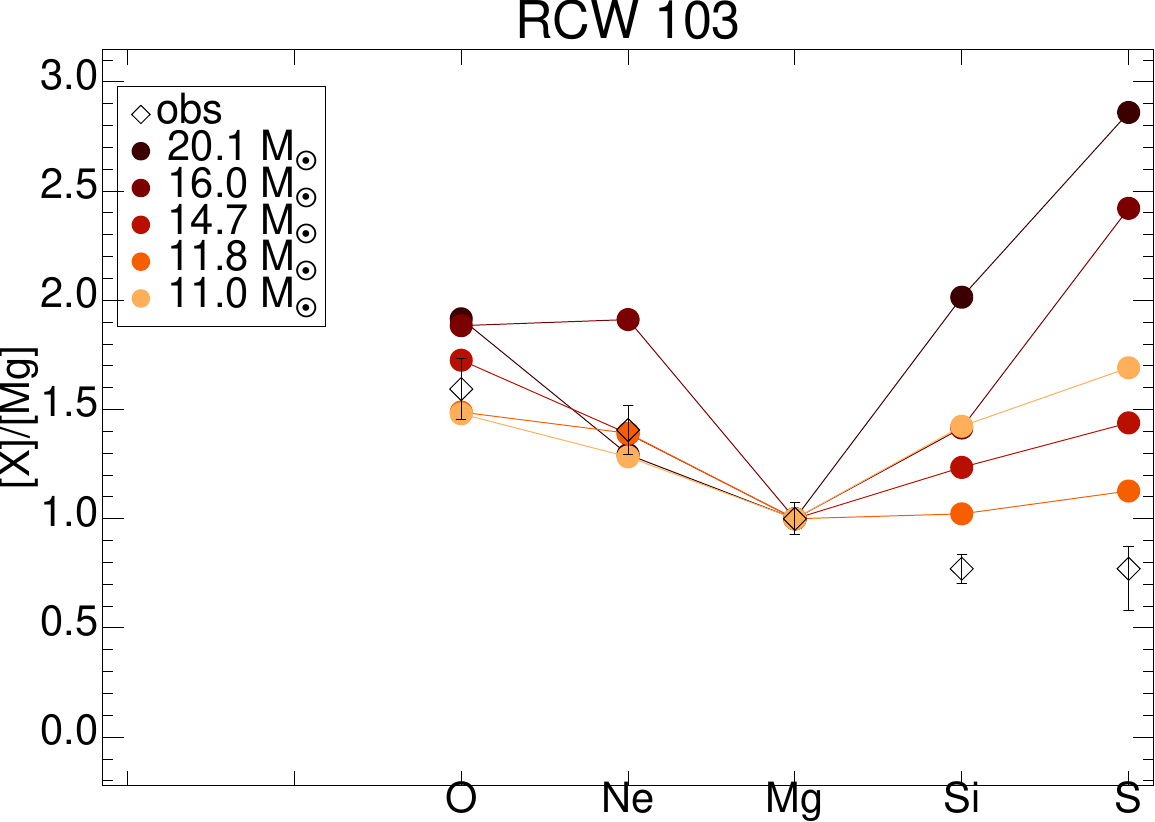}
}
{\includegraphics[angle=0, width=0.4\textwidth]{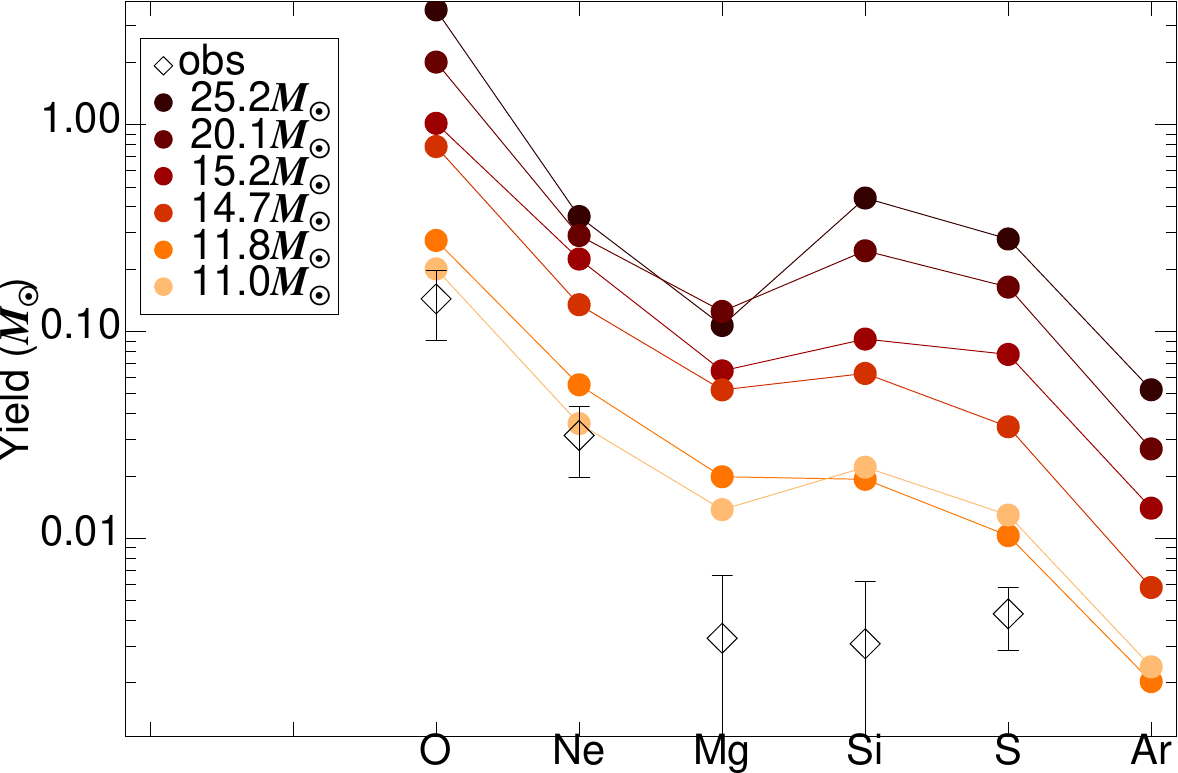}
\includegraphics[angle=0, width=0.4\textwidth]{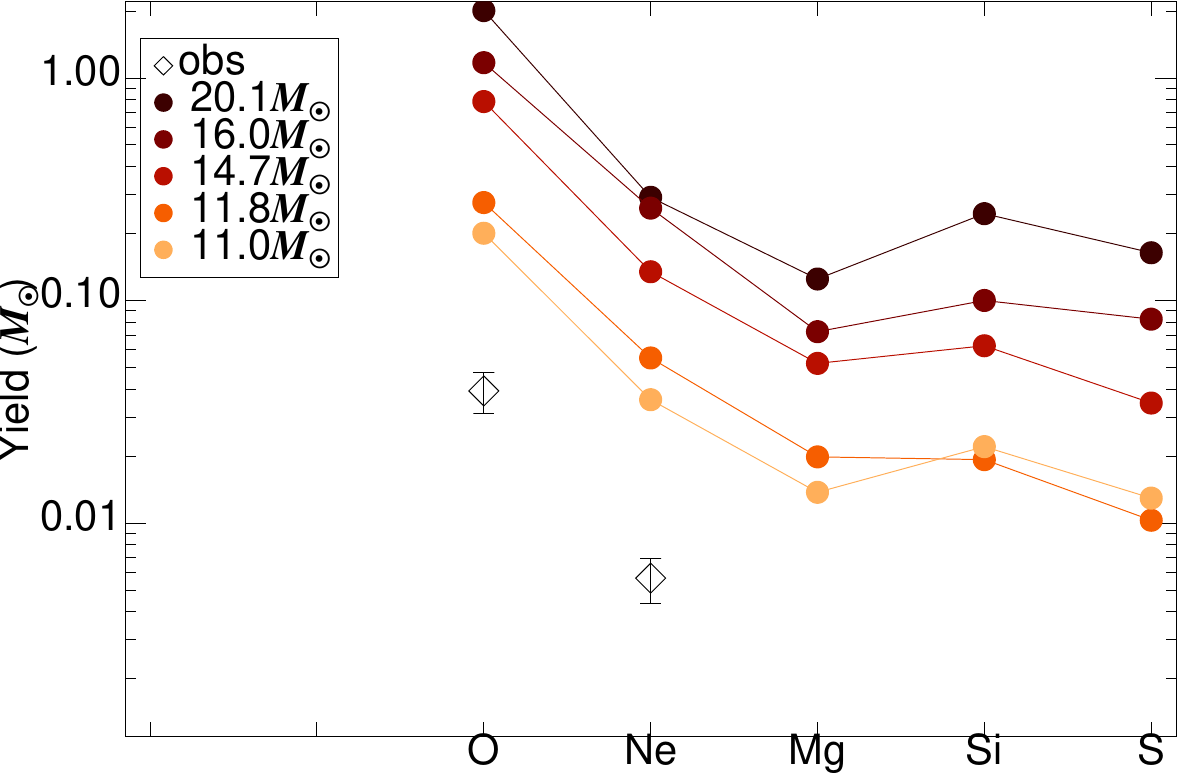}}
\caption{
Comparison of the observed abundance ratios 
(top) and metal masses (bottom) with the 
nucleosynthesis models
for a range of massive stars (dots) in \citet{sukhbold16}.
For the predicted abundances in the upper panels,
we take the shocked ISM into account by assuming
that the SN ejecta are  well mixed with the 
ISM ($M_{\rm gas}=46\Msun$ for \snra\ and 
$M_{\rm gas}=12.8\Msun$ for \snrb.)
The [Si] and [S] in \snrb\ are subsolar,
so the [Si]/[Mg] and [S]/[Mg] ratios do
not provide useful information about 
the ejecta.
}
\label{fig:promass}
\end{figure*}

The abundance ratios and masses are a useful tool to investigate 
the SN explosion, since different progenitor stars and different 
explosion mechanisms result in 
distinct ejecta patterns.
Figure~\ref{fig:ratio} shows 
the predicted abundance ratios and yields of the ejecta as a function of the 
initial masses of the progenitor stars according to the 
one-dimensional CC SN  nucleosynthesis models  \citep[solar-metallicity model W18 for stars
$> 12 \Msun$ and zero-metallicity model Z9.6 for 9--$12\Msun$ by A.\ Heger]{sukhbold16}.
\footnote{There is a large 
variation  of abundance ratios at around 20~$\Msun$. 
As stated in \citet{sukhbold16} and \citet{sukhbold14},
the transition from convective carbon core burning to radiative burning near the center at around this mass
results in highly variable pre-SN core structures and,
therefore, SN yields.}
We hereafter compare the observed abundance ratios 
and metal masses of 
the SNRs with those predicted by 
the nucleosynthesis models for CC SN explosions
(see Figs.~\ref{fig:promass} and \ref{fig:promass_n49}).

For the predicted abundances for \snra\ and
\snrb,
we take the shocked ISM into account by assuming
that the SN ejecta are mixed with the 
ISM with solar abundances.
As a result, the predicted abundance patterns 
are flatter than the pure ejecta values 
shown in Fig.~\ref{fig:ratio}.
The ratios between elements with 
(sub)solar abundances do not provide 
information for the nucleosyntheis 
models, as we cannot extract 
ejecta components.
Therefore, the Mg/Si ratio in \snra,
and Si/Mg and S/Mg ratios in \snrb\
should
not be taken seriously.

It is not always true that all the ejecta 
are mixed with the X-ray-bright ISM, 
especially for young SNRs.
Nevertheless, we take the mass-averaged abundances
across the SNRs to minimize
the problems caused by the nonuniform 
distribution of the metals.
Moreover, the observed metal masses compared
to the model values in Fig.~\ref{fig:promass} provide a clue
about the mixing level.
For \snra\ and \snrb\, the ISM masses are only
a few to ten times the typical ejecta mass.
The low metal abundances ($<2$) indicate that the total metal masses are not very large and the progenitor star is probably not very massive, as the yields are 
generally increased with the stellar mass.

\subsubsection{\snra}

By fitting the abundance ratios 
with  all the 95 models in \citet[][W18 and Z9.6, progenitor masses between 9 and 120~$\Msun$]{sukhbold16}, 
we find that the five best-fit models
for \snra\ are the 11.75 
(minimal $\chi_\nu^2$, see Fig.~\ref{fig:promass}),
14.7, 14.5, 14.9, $14.8~\Msun$ models ($\chi_\nu^2<1.4$).
The observed ejecta masses of O and Ne in
\snra\ are small: $M_{\rm O}=0.14\pm0.05~\Msun$ and 
$M_{\rm Ne}=3\pm 1\E{-2}~\Msun$.
The $\sim 11~\Msun$ model may explain the 
observed amount of O and Ne, but does not 
explain enhanced [O]/[Si] or [Ne]/[Si] ratios.
Therefore, it is possible that not 
all metals are detectable in the X-ray 
band. 
If the reverse shock has not reached the  SNR center, the inner part of
the metals could remain cold and 
the total metal masses could be underestimated.

The location of reverse shock (likely showing a 
layer of enhanced metal abundances) is not 
identified in \snra\ and \snrb.
A possible reason is that the total metal masses 
are indeed too small to emit strong X-ray
lines.
The other possibility is that the reverse shock
has already reached the SNR center.
The ratio of reverse shock radius $R_r$ to forward
shock radius $\Rs$ is related to the radial distribution of
the circumstellar medium ($n_{\rm ISM}\propto r^{-s}$) 
and ejecta $n_{\rm ejecta}\propto r^{-n}$.
For a uniform ambient medium $s=0$ and an ejecta 
power-law index $n=7$, the radius of the 
reverse shock $R_r$ can be estimated
using the solutions by \citet{truelove99} 
and an assumed ejecta 
mass of $5~\Msun$. 
In this case, the reverse shock should
have reached the SNR center.
In the $s=2$ case,  \citet{katsuda18} obtained $R_r/\Rs\sim
0.7$.

Given the large uncertainties in $R_r/\Rs$,
we consider that the progenitor mass obtained
from abundance ratios can better represent
the true values for \snra.
Nevertheless, the observed O and Ne masses 
allow us to exclude the progenitor models
with  a mass less than  $11~\Msun$.

In summary, the progenitor mass of \snra\ is
11--$15\Msun$, according to the model of
\citet{sukhbold16}. The mass of $\sim 12~\Msun$
estimated from the molecular environment (see Sect.~\ref{sec:environment}) is consistent with 
this range.
\citet{borkowski17} also obtained a relatively low mass of $\lesssim 20~\Msun$ 
by comparing  the observed metals with the nucleosynthesis model of \citet{nomoto13}.
\citet{kumar14} used two-temperature components
to fit to the X-ray data and obtained abundance ratios
overlapping ours, but they obtained a larger progenitor mass of
$\gtrsim 20~\Msun$ based on earlier nucleosynthesis
models \citep{woosley95,nomoto06} 
and the Wisconsin cross sections for the photo-electric absorption model.

\subsubsection{\snrb}

The progenitor model of $11.75~\Msun$ is
the best-fit model for O/Mg and Ne/Mg ratios 
in \snrb\ (see Fig.~\ref{fig:promass}), and  the five best models are 11.75, 17.6, 14.7, 12.0, and $17.4~\Msun$
models.
Here the Si/Mg and S/Mg ratios are 
not fitted.

The explosion energy of \snrb\ 
($\sim 10^{50}~\erg$) is among the weakest 
in Galactic SNRs. 
Among the five well-fit models, two 
have progenitor masses $\le 12~\Msun$
and relatively weak explosion energy 
($E_0=2.6$--$6.6\E{50}~\erg$),
while the other models
have a canonical explosion energy.

The total plasma mass of $\sim 13~\Msun$
is only a few times larger than the expected
ejecta mass for a CCSN from a normal 
explosion ($\gtrsim 5~\Msun$).
If all the ejecta have been heated by the shocks, 
we would expect to see high metal abundances.
One possibility is that most of the ejecta are cool 
and not probed in the X-ray band for the normal
SN explosion scenario.
An alternative explanation is that the ejecta mass
is indeed small because of significant fallback
from a weak CCSN explosion (see discussion below).

The overall [Si] and [S] are subsolar,
suggesting that overall Si and S 
production is low in \snrb.
Although a few Si/S-rich bins are 
detected in the SNRs (see Fig.~\ref{fig:rcw103_pars}),
they  may correspond to some pure ejecta knots
\citep[see also][for Si and S ejecta knots]{frank15}. 
The distribution of [Mg] gives a clue 
to the missing Si/S problem. 
As shown in Figs.~\ref{fig:rcw103_pars} and 
\ref{fig:profile}, the Mg element is oversolar
in the SNR interior, but decreases to subsolar
in the outer region.
This implies that the heavier elements are
distributed more in the inner regions compared
to the lighter elements such as O and Ne.
The Si and S materials may have smaller 
ejection velocities and the layers might
not be heated by the reverse shock.
A more extreme case is that the elements heavier
than Mg may fall back to the compact objects
due to a weak SN explosion.

The weak explosion energy of \snrb\ means that
the total ejecta mass or the initial velocity
of the ejecta should be smaller than normal 
SNRs. 
According to the simulation of a CCSN explosion 
invoking convective engine \citep{fryer18},
a weaker SN explosion results in a more massive
neutron star and less ejecta due to fallback. 
Their simulation considered 15, 20, and 25~$\Msun$ cases. The weakest
explosion ($3.4\E{50}~\erg$) of a $15~\Msun$ star creates a $1.9~\Msun$
compact remnant, and more productions of
O ($0.29~\Msun$)  and Ne ($0.064~\Msun$) 
than observed in \snrb.
For a star more massive, a weak
explosion would create a black hole.
Therefore, we suggest that the
progenitor star of \snrb\ has a 
mass of $\lesssim 13~\Msun$, based on
a comparison 
with the nucleosynthesis  models,
and the fact that  the existence of nearby 
molecular shells disfavor a star more 
massive than $13~\Msun$ (see Sect.~\ref{sec:environment}).
The two-temperature analysis of \snrb\ leads
to a comparable progenitor mass and low explosion 
energy \citep{braun19}.
However, the progenitor mass derived here is 
lower than the value 
of 18--20~$\Msun$ obtained from  \citet{frank15} 
using an earlier nucleosynthesis model \citep{nomoto06}.

The low explosion energy, the small observed
metal masses, and low abundances of heavier elements such as Si and S,
consistently suggest 
that \snrb\ is produced 
by a weak SN explosion with significant fallback. 
It has been suggested that a supernova 
fallback disk may be a critical ingredient 
in explaining the very long spin 
period of 1E~161348$-$5055 in 
\snrb\ \citep{deluca06,li07,tong16,rea16,xu19}.
Our study supports this fallback
scenario.
In this case, the significant amount 
of fallback materials increase the mass of the compact object.
Therefore, we predict that 1E~161348$-$5055
is a relatively massive neutron star.

\subsubsection{N49}

\begin{figure}
  \centering
\includegraphics[angle=0, width=0.45\textwidth]{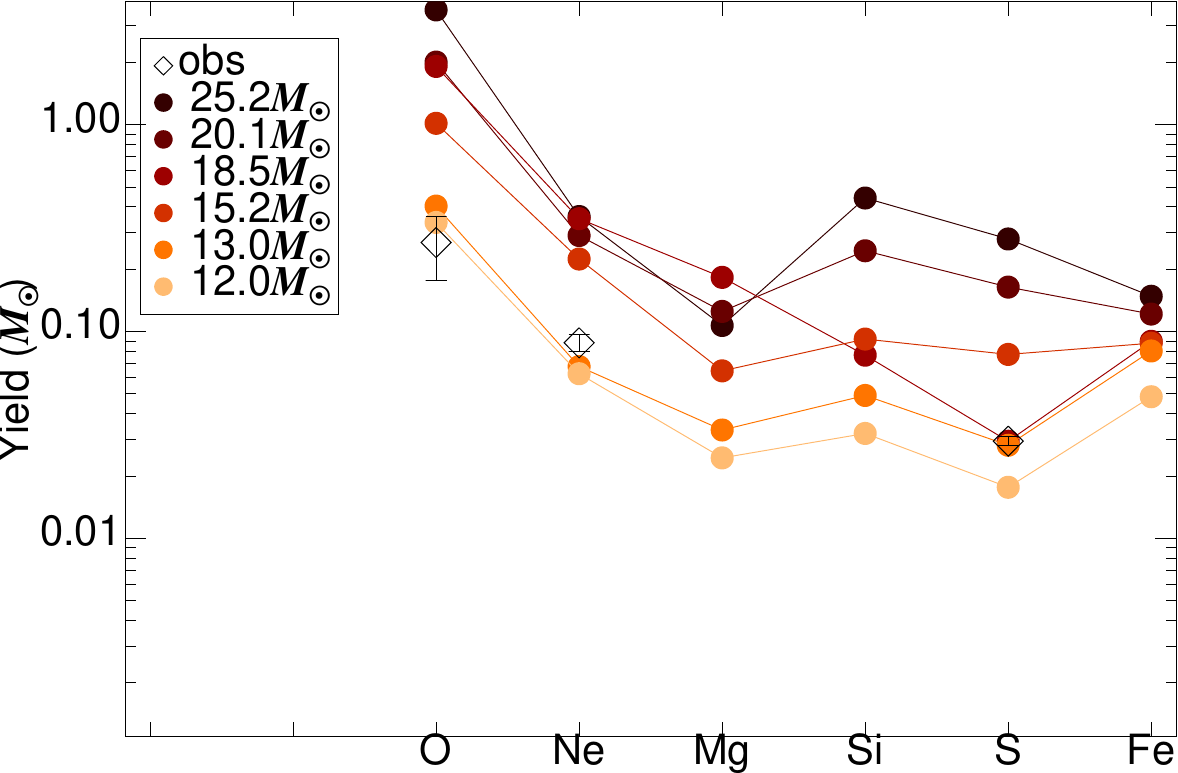}
\caption{
Comparison of 
the metal masses in \snrc\ with the 
nucleosynthesis models of \citet{sukhbold16}.
The masses of Mg, Si, and Fe
are not plotted because they are
lower than the typical values in the LMC.
}
\label{fig:promass_n49}
\end{figure}

\snrc\ is located in the LMC, while
the W18 models are applied for stars
with solar metallicity. 
Nevertheless, the lower metallicity mainly influences
the mass loss of the stars but 
affects less the
evolution in the core; therefore, 
the overall results from the core may be 
similar to those with solar metallicity
for stars below 30~$\Msun$ (private 
communication with T. Sukhbold).

The measured Si abundance of $\sim 0.6$
is clearly lower than the typical value of 0.87
in the LMC \citep{hanke10}; this means the
uncertainties of abundance ratios could
be larger than the measurement values given
the variation of the LMC abundance.
Therefore,  we only show a comparison of the 
measured metal  masses with the 
nucleosynthesis model in Fig.~\ref{fig:promass_n49}.
The $13~\Msun$ model gives a relatively
good fit to observed metal masses of O, Ne,
and S. This puts a lower limit on the 
progenitor mass of \snrc, as lower mass stars produce less of these metals.
The nucleosynthesis models predict that
the 15--17$~\Msun$ stars 
produce abundance
patterns with enhanced O and Ne relative to Si, and also 
enhanced S relative to O 
(see Fig.~\ref{fig:ratio}),
which is the case for \snrc.
Although a $\sim 26~\Msun$ star may also produce
these abundance patterns, 
it is not very likely to be the progenitor of \snrc\  as 
its SN yields would be over one order of magnitude larger than
the observed metal masses.
Therefore,  it is likely that \snrc\ has a progenitor
with a mass between 13 and 17~$\Msun$.
This is consistent with the suggestion
that \snrc\ has an early B-type progenitor
\citep{schull85},
while the  progenitor mass obtained by
\cite{park03b}
is larger ($\gtrsim 25~\Msun$) based on 
enhanced Mg (not as enhanced here) and a
comparison with an earlier nucleosynthesis
model  \citep{thielemann96}.

\subsection{Implication for the formation of the magnetars}

The progenitors of the three magnetars 
have the stellar masses of $< 20~\Msun$ 
(11--$15~\Msun$ for \snra, $\lesssim 13~\Msun$ for \snrb\ and $\sim 13$--$17~\Msun$ for \snrc),
consistent with B type stars rather
than more massive O type stars.
The relatively low-mass progenitor stars 
of these three SNRs are also supported by \citet[$M_{\rm ZAMS}<15~\Msun$]{katsuda18b},
based on elemental 
abundances 
in the literature.
Therefore, magnetars are not 
all necessarily made from very massive stars.

While 
there is so far 
no consensus on magnetar progenitors, there is
evidence that some of them originate from very massive progenitors  \citep[$M_{\rm ZAMS}>30~\Msun$, see][and references therein]{safiharb13}. 
A piece of evidence for very massive progenitors 
comes from  the study of the magnetar CXO~J164710.2$-$455216 in the massive star cluster Westerlund~1. The age and the stars of the stellar cluster suggest that 
the progenitor star of this magnetar
has an initial mass of over 40~$\Msun$
\citep{muno06}.
However, \citet{aghakhanloo19}
reduced the distance of the cluster
from 5~kpc to $\sim 3.2$~kpc
using Gaia data release 2 parallaxes.
This revises the progenitor mass 
of the magnetar to $25~\Msun$.
Another magnetar, SGR~1806$-$20,  in a massive star cluster was likely created by a star with a mass greater
than $50~\Msun$ \citep{figer05}.
On the other hand, there is evidence that 
magnetars and high magnetic field pulsars are from lower
mass stars, in addition to the three magnetars
studied here.
The SNR~Kes 75 that hosts
a high magnetic field pulsar J1846$-$0258 
\citep[but shows magnetar-like bursts,][]{gavriil08} was considered to have a Wolf-Rayet
progenitor \citep{morton07}.
However,  the existence of a molecular shell surrounding 
it suggests a progenitor mass of 
$12\pm2 \Msun$ for Kes~75 \citep{chen13}. 
A similar low mass (8--$12~\Msun$) 
was obtained from the far-IR observations and
a comparison to the 
nucleosynthesis models \citep{temim19}. 
Moreover, the magnetar SGR~1900+14 in
a stellar cluster is suggested to have a progenitor mass of $17\pm 2~\Msun$ \citep{davies09}.

Magnetars are likely made
from stars that span a large mass
range. 
According to current knowledge about
Galactic magnetars with progenitor information, most magnetars, though not all of them, seem 
to result from stars with $<20~\Msun$.
Among the three magnetars in our study,
\snrc\  seems to have a higher
progenitor mass (13--$17~\Msun$) 
than \snrb\ ($\lesssim 13~\Msun$).

The SN explosion energies of the three magnetars are not very 
high, ranging from $10^{50}~\erg$ to 
$\sim 1.7\E{51}~\erg$, 
supporting the possibility that their SN explosions were not 
significantly powered by rapidly spinning magnetars.
Particularly, \snrb, 
the remnant hosting an ultra-slow magnetar with 
a rotational period of $P=6.67$~hr, resulted from a weak 
explosion with energy an order of magnitude lower
than the canonical value.
The SNR CTB~37B that hosts CXOU J171405.7$-$381031 also 
resulted from a weak explosion
\citep[$1.8\pm0.6\E{50}~\erg$,][]{blumer19}.
Furthermore, CTB~109 (hosting 1E~2259+586) has a 
normal \citep{sasaki04,vink06c} or even low explosion 
energy \citep[2--$5\E{50}~\erg$, see][for a recent measurement and see references therein]{sanchez-cruces18}.
The low-to-normal SN explosion energy appears to be a
common property of the known magnetar-SNR systems
with extended thermal X-ray emission. 

As pointed out in an earlier paper by \citet{vink06c},
the relatively low or canonical explosion energy
does not suggest that these three magnetars were 
born with very rapidly spinning millisecond 
pulsars.
The rotational energy of a neutron star is
$E_{\rm rot}\approx 3\E{52} (P/{\rm 1~ms})^{-2}
\erg$.
A rapidly spinning magnetar loses its rotational energy quickly
\citep[$\sim 10$--100 s,][]{thompson04}. 
During the first few weeks, the magnetar energy 
goes into accelerating and heating the ejecta as the SN is
optically thick, and at a later stage, the energy is released 
through radiation \citep{woosley10}. 
This suggests that millisecond magnetars can lose some of their
energies to the SN kinetic energies \citep[$\sim 40\%$ in the model by][but this fraction could be highly uncertain]{woosley10}.
\citet{dallosso09} proposed that gravitational waves might
also take away the magnetar energy.
The quickly rotating millisecond magnetar is regarded as 
a likely central engine for Type I 
superluminous supernovae
\citep[e.g.,][]{woosley10,kasen10}.
According to both theoretical studies and observations,
superluminous SNe powered by millisecond magnetars should have 
significantly enhanced  kinetic energies \citep[2--$10\E{51}~\erg$,][]{nicholl17,soker17}.
The three SNRs studied in this paper, in addition to CTB~109 and
CTB~37B, have much lower kinetic
energies than those of Type I superluminous SNe, 
indicating that their origin is different.

The distribution of the metals reveals some asymmetries:
\snra\ likely has enhanced O, Ne, and Mg 
abundances in the east 
(but could
also be a result of the degeneracy
between [O] and $\NH$ in spectral
fit),
\snrb\ shows enhanced O abundance
in the south, and \snrc\ shows
clearly enhanced O, Ne, and Mg, Si
toward the east, and enhanced
S in the south.
The element distributions are not always anti-correlated
with the density, so the dilution
due to the ejecta--gas mixing 
cannot  be the main reason for 
the  observed asymmetries, 
especially for \snra\ and \snrb.
The nonuniform ejecta 
distributions reflect
that the SN  explosions should 
be aspherical to some extent.

With the above information, we can distinguish
the two hypotheses about the origin of magnetars:
dynamo origin or fossil field origin.
The dynamo model predicts that
the SN explosion is energized by the millisecond pulsar,
which has been ruled out for the 
three magnetars discussed in this paper.
Furthermore, the rapidly rotating stars are 
generally made from  very massive stars 
\citep[$\le 3$ ms pulsars from $35~\Msun$ stars,][]{heger05}.
The SN rate is $\lesssim 5\%$ for stars with an initial
mass $>30~\Msun$ and $\sim 10\%$ for stars $>20\Msun$
\citep{sukhbold16}.
These very massive stars are suggested to
collapse to form black holes rather than neutron stars \citep{fryer99,smartt09}.
Therefore, it is likely that only a 
small fraction of magnetars may be 
formed through this  dynamo channel. 
We obtain a normal mass range 
($M_{\rm ZAMS}<20~\Msun$) for the progenitor stars 
of the three magetars, 
further disfavoring the dynamo scenario for them.

The fossil field origin appears to be a natural 
explanation for magnetars.
The magnetic field strengths of
massive stars vary by a few orders
of magnitude.
The magnetic field detection rate is
$\sim 7\%$ for both B-type and O-type stars, with 
magnetic fields from several hundred~Gauss
to over 10 kG
\citep[e.g.,][]{grunhut12,wade14,scholler17}.
As to the origin of the strong magnetic 
fields in magnetic stars, the debates 
are almost the same as for magnetars:
dynamo or fossil.
The latter origin is supported from 
both theoretical studies and observations
in recent years.
Theoretical study of magnetic stars and 
magnetars has shown  that stable, twisted 
magnetic fields (poloidal fields above the surface + 
internal toroidal fields) can have evolved from random initial fields
\citep{braithwaite04, braithwaite09}.
Recent observations confirmed
the fossil field origin \citep{neiner15}, because
the dynamo origin would lead 
to a correlation between the magnetic
field strength and stellar rotation speed, 
which is not observed.
It is even suggested that the massive stars with 
higher magnetic fields rotate more slowly, likely due to magnetic braking \citep{shultz18}.
Fossil magnetic fields of the stars
are descendants from the seed fields
of the parent molecular clouds
\citep{mestel99}. 
After the death of the stars, the neutron stars 
may also inherit the magnetic fields from these stars.

In our Galaxy, ten magnetars have been found in SNRs.
Among the 295--383  known Galactic SNRs \citep{ferrand12,green14,green17},
around 80\%  are of CC origin 
\citep[$0.81\pm 0.24$,][]{li11}.
This means that $\sim 3\%$-- $4\%$ of CCSNRs 
are found to host magnetars. 
This fraction is slightly smaller 
than the incident fraction of magnetic 
OB stars with magnetic fields over a few 
hundred Gauss ($\sim 7\%$), but consistent
with the fraction of massive stars with higher fields 
\citep[$\sim 3\%$ with $B >10^3\G$,][]{scholler17}.
Therefore, our study supports the fossil 
field origin as an important channel
to produce magnetars, given 
the  normal mass range ($M_{\rm ZAMS} < 20 \Msun$) of
the progenitor stars, 
the low-to-normal explosion energy of 
the SNRs,  and the fraction
of magnetars found in SNRs.
Although our current study favors the fossil
field origin and is against the dynamo origin for
the three magnetars, we do not
exclude the possibility that there 
might be more
than one channel to create magnetars.

\section{Conclusions}

We have performed a spatially resolved X-ray study
of SNRs \snra, \snrb, and \snrc, aiming
to learn how their magnetars are formed.
Our study supports the fossil field as a probable origin of 
these magnetars.
The main conclusions are summarized as follows:

\begin{enumerate}
   \item 
   The progenitor stars of the three magnetars are
   $< 20~\Msun$ (11--$15~\Msun$ for \snra, $\lesssim 13~\Msun$ for \snrb,\ and $\sim 13$--$17~\Msun$ for \snrc). 
   The progenitor masses are obtained
   using two methods: 1) a comparison of the metal abundances and masses with supernova nucleosynthesis models,
   2) the nearby molecular shell that cannot be
   explained with very massive progenitor stars, which can launch strong main-sequence winds
   and create large low-density cavities.
   The two methods give consistent results.
   
   \item Magnetars are likely made from stars  that span a large mass range. According to current knowledge about
Galactic magnetars, 
a good fraction of the magnetars
seems  to result from stars $<20~\Msun$.

    \item 
    The explosion energies of the three SNRs
    span a large range, from $10^{50}~\erg$ to  
    $\sim 1.7\E{51}~\erg$, further
    supporting  that  their  SN  explosions  had  not  been  significantly powered by millisecond magnetars.

   \item
   A common characteristic among the three SNRs 
   is that all of them are O- and Ne-enhanced, and there 
   is no evidence of overabundant Fe (average value). 
   The distribution  of  the  metals  reveals  some  asymmetries, reflecting that the SN explosions are probably aspherical to some extent.
   The next common property is that they are
   likely to be interacting with molecular structures swept up
   by winds of the progenitor stars.
   
   \item
    We report that \snrb\  is produced by a weak SN
    explosion  with  significant fallback, given  
    the low explosion energy ($\sim 10^{50}\dub^{2.5}~\erg$), the small observed metal masses 
    ($M_{\rm O}\sim 4\E{-2}~\Msun$ and $M_{\rm Ne}\sim 6\E{-3}~\Msun$), and sub-solar abundances of 
    heavier elements such as Si and S. 
    This supports the fallback scenario in explaining 
    the very  long spin period of 1E 161348$-$5055.

    \item 
    Our study supports the fossil field as a probable origin of many magnetars.
    The dynamo scenario, involving rapid
    initial spinning of the neutron star, is not supported
    given 
    the normal mass range ($M_{\rm ZAMS} < 20\Msun$) 
    of the progenitor stars,  
    the low-to-normal explosion energy of the 
    SNRs, 
    and the large number of known
    Galactic magnetars. 
    On the contrary, the fraction of CCSNRs hosting magnetars is consistent with the fraction of magnetic OB stars with high fields.

\end{enumerate}


\begin{acknowledgements}
We thank Tuguldur Sukhbold for valuable discussions
about his SN nucleosynthesis models.
We also thank Hao Tong for helpful comments on the origin 
of magnetars.
P.Z. acknowledges the support from the NWO Veni Fellowship, grant no. 639.041.647
and NSFC grants 11503008 and 11590781.
S.S.H. acknowledges support from the Natural Sciences and Engineering Research Council of Canada (NSERC).
\end{acknowledgements}



\begin{thebibliography}{90}
\expandafter\ifx\csname natexlab\endcsname\relax\def\natexlab#1{#1}\fi

\bibitem[{{Aghakhanloo} {et~al.}(2019){Aghakhanloo}, {Murphy}, {Smith},
  {Parejko}, {D{\'\i}az-Rodr{\'\i}guez}, {Drout}, {Groh}, {Guzman}, \&
  {Stassun}}]{aghakhanloo19}
{Aghakhanloo}, M., {Murphy}, J.~W., {Smith}, N., {et~al.} 2019, arXiv e-prints,
  arXiv:1901.06582

\bibitem[{{Asplund} {et~al.}(2009){Asplund}, {Grevesse}, {Sauval}, \&
  {Scott}}]{asplund09}
{Asplund}, M., {Grevesse}, N., {Sauval}, A.~J., \& {Scott}, P. 2009, \araa, 47,
  481

\bibitem[{{Banas} {et~al.}(1997){Banas}, {Hughes}, {Bronfman}, \&
  {Nyman}}]{banas97}
{Banas}, K.~R., {Hughes}, J.~P., {Bronfman}, L., \& {Nyman}, L.~{\r{A}}. 1997,
  \apj, 480, 607

\bibitem[{{Blumer} {et~al.}(2019){Blumer}, {Safi-Harb}, {Kothes}, {Rogers}, \&
  {Gotthelf}}]{blumer19}
{Blumer}, H., {Safi-Harb}, S., {Kothes}, R., {Rogers}, A., \& {Gotthelf}, E.~V.
  2019, \mnras, 1572

\bibitem[{{Borkowski} \& {Reynolds}(2017)}]{borkowski17}
{Borkowski}, K.~J. \& {Reynolds}, S.~P. 2017, \apj, 846, 13

\bibitem[{{Borkowski} {et~al.}(2001){Borkowski}, {Rho}, {Reynolds}, \&
  {Dyer}}]{borkowski01}
{Borkowski}, K.~J., {Rho}, J., {Reynolds}, S.~P., \& {Dyer}, K.~K. 2001, \apj,
  550, 334

\bibitem[{{Braithwaite}(2009)}]{braithwaite09}
{Braithwaite}, J. 2009, \mnras, 397, 763

\bibitem[{{Braithwaite} \& {Spruit}(2004)}]{braithwaite04}
{Braithwaite}, J. \& {Spruit}, H.~C. 2004, \nat, 431, 819

\bibitem[{{Braun} {et al.}(2019)}]{braun19}
Braun, C., Safi-Harb, S., \& Fryer, C. 2019, MNRAS, accepted

\bibitem[{{Cappellari} \& {Copin}(2003)}]{cappellari03}
{Cappellari}, M. \& {Copin}, Y. 2003, \mnras, 342, 345

\bibitem[{{Carter} {et~al.}(1997){Carter}, {Dickel}, \& {Bomans}}]{carter97}
{Carter}, L.~M., {Dickel}, J.~R., \& {Bomans}, D.~J. 1997, \pasp, 109, 990

\bibitem[{{Chen} {et~al.}(2013){Chen}, {Zhou}, \& {Chu}}]{chen13}
{Chen}, Y., {Zhou}, P., \& {Chu}, Y.-H. 2013, \apjl, 769, L16

\bibitem[{{Chevalier}(1999)}]{chevalier99}
{Chevalier}, R.~A. 1999, \apj, 511, 798

\bibitem[{{Chevalier}(2005)}]{chevalier05}
{Chevalier}, R.~A. 2005, \apj, 619, 839

\bibitem[{{D'A{\`\i}} {et~al.}(2016){D'A{\`\i}}, {Evans}, {Burrows}, {Kuin},
  {Kann}, {Campana}, {Maselli}, {Romano}, {Cusumano}, {La Parola}, {Barthelmy},
  {Beardmore}, {Cenko}, {De Pasquale}, {Gehrels}, {Greiner}, {Kennea}, {Klose},
  {Melandri}, {Nousek}, {Osborne}, {Palmer}, {Sbarufatti}, {Schady}, {Siegel},
  {Tagliaferri}, {Yates}, \& {Zane}}]{dai16}
{D'A{\`\i}}, A., {Evans}, P.~A., {Burrows}, D.~N., {et~al.} 2016, \mnras, 463,
  2394

\bibitem[{{Dall'Osso} {et~al.}(2009){Dall'Osso}, {Shore}, \&
  {Stella}}]{dallosso09}
{Dall'Osso}, S., {Shore}, S.~N., \& {Stella}, L. 2009, \mnras, 398, 1869

\bibitem[{{Davies} {et~al.}(2009){Davies}, {Figer}, {Kudritzki}, {Trombley},
  {Kouveliotou}, \& {Wachter}}]{davies09}
{Davies}, B., {Figer}, D.~F., {Kudritzki}, R.-P., {et~al.} 2009, \apj, 707, 844

\bibitem[{{De Luca} {et~al.}(2006){De Luca}, {Caraveo}, {Mereghetti}, {Tiengo},
  \& {Bignami}}]{deluca06}
{De Luca}, A., {Caraveo}, P.~A., {Mereghetti}, S., {Tiengo}, A., \& {Bignami},
  G.~F. 2006, Science, 313, 814

\bibitem[{{Diehl} \& {Statler}(2006)}]{diehl06}
{Diehl}, S. \& {Statler}, T.~S. 2006, \mnras, 368, 497

\bibitem[{{Ferrand} \& {Safi-Harb}(2012)}]{ferrand12}
{Ferrand}, G. \& {Safi-Harb}, S. 2012, Advances in Space Research, 49, 1313

\bibitem[{{Ferrario} \& {Wickramasinghe}(2006)}]{ferrario06}
{Ferrario}, L. \& {Wickramasinghe}, D. 2006, \mnras, 367, 1323

\bibitem[{{Ferrario} \& {Wickramasinghe}(2008)}]{ferrario08}
{Ferrario}, L. \& {Wickramasinghe}, D. 2008, \mnras, 389, L66

\bibitem[{{Figer} {et~al.}(2005){Figer}, {Najarro}, {Geballe}, {Blum}, \&
  {Kudritzki}}]{figer05}
{Figer}, D.~F., {Najarro}, F., {Geballe}, T.~R., {Blum}, R.~D., \& {Kudritzki},
  R.~P. 2005, \apjl, 622, L49

\bibitem[{{Frank} {et~al.}(2015){Frank}, {Burrows}, \& {Park}}]{frank15}
{Frank}, K.~A., {Burrows}, D.~N., \& {Park}, S. 2015, \apj, 810, 113

\bibitem[{{Fryer}(1999)}]{fryer99}
{Fryer}, C.~L. 1999, \apj, 522, 413

\bibitem[{{Fryer} {et~al.}(2018){Fryer}, {Andrews}, {Even}, {Heger}, \&
  {Safi-Harb}}]{fryer18}
{Fryer}, C.~L., {Andrews}, S., {Even}, W., {Heger}, A., \& {Safi-Harb}, S.
  2018, \apj, 856, 63

\bibitem[{{Gavriil} {et~al.}(2008){Gavriil}, {Gonzalez}, {Gotthelf}, {Kaspi},
  {Livingstone}, \& {Woods}}]{gavriil08}
{Gavriil}, F.~P., {Gonzalez}, M.~E., {Gotthelf}, E.~V., {et~al.} 2008, Science,
  319, 1802

\bibitem[{{Green}(2014)}]{green14}
{Green}, D.~A. 2014, Bulletin of the Astronomical Society of India, 42, 47

\bibitem[{{Green}(2017)}]{green17}
{Green}, D.~A. 2017, VizieR Online Data Catalog, 7278

\bibitem[{{Grevesse} \& {Sauval}(1998)}]{grevesse98}
{Grevesse}, N. \& {Sauval}, A.~J. 1998, \ssr, 85, 161

\bibitem[{{Grunhut} {et~al.}(2012){Grunhut}, {Wade}, \& {MiMeS
  Collaboration}}]{grunhut12}
{Grunhut}, J.~H., {Wade}, G.~A., \& {MiMeS Collaboration}. 2012, in
  Astronomical Society of the Pacific Conference Series, Vol. 465, Proceedings
  of a Scientific Meeting in Honor of Anthony F. J. Moffat, ed. L.~{Drissen},
  C.~{Robert}, N.~{St-Louis}, \& A.~F.~J. {Moffat}, 42

\bibitem[{{Hanke} {et~al.}(2010){Hanke}, {Wilms}, {Nowak}, {Barrag{\'a}n}, \&
  {Schulz}}]{hanke10}
{Hanke}, M., {Wilms}, J., {Nowak}, M.~A., {Barrag{\'a}n}, L., \& {Schulz},
  N.~S. 2010, \aap, 509, L8

\bibitem[{{Heger} {et~al.}(2005){Heger}, {Woosley}, \& {Spruit}}]{heger05}
{Heger}, A., {Woosley}, S.~E., \& {Spruit}, H.~C. 2005, \apj, 626, 350

\bibitem[{{Hu} \& {Lou}(2009)}]{hu09}
{Hu}, R.-Y. \& {Lou}, Y.-Q. 2009, \mnras, 396, 878

\bibitem[{{Kasen} \& {Bildsten}(2010)}]{kasen10}
{Kasen}, D. \& {Bildsten}, L. 2010, \apj, 717, 245

\bibitem[{{Kaspi} \& {Beloborodov}(2017)}]{kaspi17}
{Kaspi}, V.~M. \& {Beloborodov}, A.~M. 2017, \araa, 55, 261

\bibitem[{{Katsuda} {et~al.}(2018{\natexlab{a}}){Katsuda}, {Morii}, {Janka},
  {Wongwathanarat}, {Nakamura}, {Kotake}, {Mori}, {M{\"u}ller}, {Takiwaki},
  {Tanaka}, {Tominaga}, \& {Tsunemi}}]{katsuda18}
{Katsuda}, S., {Morii}, M., {Janka}, H.-T., {et~al.} 2018{\natexlab{a}}, \apj,
  856, 18

\bibitem[{{Katsuda} {et~al.}(2018{\natexlab{b}}){Katsuda}, {Takiwaki},
  {Tominaga}, {Moriya}, \& {Nakamura}}]{katsuda18b}
{Katsuda}, S., {Takiwaki}, T., {Tominaga}, N., {Moriya}, T.~J., \& {Nakamura},
  K. 2018{\natexlab{b}}, \apj, 863, 127

\bibitem[{{Kumar} {et~al.}(2014){Kumar}, {Safi-Harb}, {Slane}, \&
  {Gotthelf}}]{kumar14}
{Kumar}, H.~S., {Safi-Harb}, S., {Slane}, P.~O., \& {Gotthelf}, E.~V. 2014,
  \apj, 781, 41

\bibitem[{{Li} {et~al.}(2011){Li}, {Chornock}, {Leaman}, {Filippenko},
  {Poznanski}, {Wang}, {Ganeshalingam}, \& {Mannucci}}]{li11}
{Li}, W., {Chornock}, R., {Leaman}, J., {et~al.} 2011, \mnras, 412, 1473

\bibitem[{{Li}(2007)}]{li07}
{Li}, X.-D. 2007, \apj, 666, L81

\bibitem[{{Liu} {et~al.}(2017){Liu}, {Chen}, {Zhang}, {Liu}, {He}, {Zhou},
  {Zhou}, \& {Su}}]{liu17}
{Liu}, B., {Chen}, Y., {Zhang}, X., {et~al.} 2017, \apj, 851, 37

\bibitem[{{Martin} {et~al.}(2014){Martin}, {Rea}, {Torres}, \&
  {Papitto}}]{martin14}
{Martin}, J., {Rea}, N., {Torres}, D.~F., \& {Papitto}, A. 2014, \mnras, 444,
  2910

\bibitem[{{Mestel}(1999)}]{mestel99}
{Mestel}, L. 1999, {Stellar magnetism}

\bibitem[{{Morton} {et~al.}(2007){Morton}, {Slane}, {Borkowski}, {Reynolds},
  {Helfand}, {Gaensler}, \& {Hughes}}]{morton07}
{Morton}, T.~D., {Slane}, P., {Borkowski}, K.~J., {et~al.} 2007, \apj, 667, 219

\bibitem[{{Muno} {et~al.}(2006){Muno}, {Clark}, {Crowther}, {Dougherty}, {de
  Grijs}, {Law}, {McMillan}, {Morris}, {Negueruela}, {Pooley}, {Portegies
  Zwart}, \& {Yusef-Zadeh}}]{muno06}
{Muno}, M.~P., {Clark}, J.~S., {Crowther}, P.~A., {et~al.} 2006, \apjl, 636,
  L41

\bibitem[{{Nakamura} {et~al.}(2009){Nakamura}, {Bamba}, {Ishida}, {Nakajima},
  {Yamazaki}, {Terada}, {P{\"u}hlhofer}, \& {Wagner}}]{nakamura09}
{Nakamura}, R., {Bamba}, A., {Ishida}, M., {et~al.} 2009, \pasj, 61, S197

\bibitem[{{Neiner} {et~al.}(2015){Neiner}, {Mathis}, {Alecian}, {Emeriau},
  {Grunhut}, {BinaMIcS}, \& {MiMeS Collaborations}}]{neiner15}
{Neiner}, C., {Mathis}, S., {Alecian}, E., {et~al.} 2015, in IAU Symposium,
  Vol. 305, Polarimetry, ed. K.~N. {Nagendra}, S.~{Bagnulo}, R.~{Centeno}, \&
  M.~{Jes{\'u}s Mart{\'{\i}}nez Gonz{\'a}lez}, 61--66

\bibitem[{{Nicholl} {et~al.}(2017){Nicholl}, {Guillochon}, \&
  {Berger}}]{nicholl17}
{Nicholl}, M., {Guillochon}, J., \& {Berger}, E. 2017, \apj, 850, 55

\bibitem[{{Nomoto} {et~al.}(2013){Nomoto}, {Kobayashi}, \&
  {Tominaga}}]{nomoto13}
{Nomoto}, K., {Kobayashi}, C., \& {Tominaga}, N. 2013, \araa, 51, 457

\bibitem[{{Nomoto} {et~al.}(2006){Nomoto}, {Tominaga}, {Umeda}, {Kobayashi}, \&
  {Maeda}}]{nomoto06}
{Nomoto}, K., {Tominaga}, N., {Umeda}, H., {Kobayashi}, C., \& {Maeda}, K.
  2006, Nuclear Physics A, 777, 424

\bibitem[{{Olausen} \& {Kaspi}(2014)}]{olausen14}
{Olausen}, S.~A. \& {Kaspi}, V.~M. 2014, \apjs, 212, 6

\bibitem[{{Oliva} {et~al.}(1999){Oliva}, {Moorwood}, {Drapatz}, {Lutz}, \&
  {Sturm}}]{oliva99}
{Oliva}, E., {Moorwood}, A.~F.~M., {Drapatz}, S., {Lutz}, D., \& {Sturm}, E.
  1999, \aap, 343, 943

\bibitem[{Ostriker \& McKee(1988)}]{ostriker88}
Ostriker, J.~P. \& McKee, C.~F. 1988, Rev. Mod. Phys., 60, 1

\bibitem[{{Otsuka} {et~al.}(2010){Otsuka}, {van Loon}, {Long}, {Meixner},
  {Matsuura}, {Reach}, {Roman-Duval}, {Gordon}, {Sauvage}, {Hony}, {Misselt},
  {Engelbracht}, {Panuzzo}, {Okumura}, {Woods}, {Kemper}, \&
  {Sloan}}]{otsuka10}
{Otsuka}, M., {van Loon}, J.~T., {Long}, K.~S., {et~al.} 2010, \aap, 518, L139

\bibitem[{{Park} {et~al.}(2003){Park}, {Hughes}, {Burrows}, {Slane}, {Nousek},
  \& {Garmire}}]{park03b}
{Park}, S., {Hughes}, J.~P., {Burrows}, D.~N., {et~al.} 2003, \apjl, 598, L95

\bibitem[{{Park} {et~al.}(2012){Park}, {Hughes}, {Slane}, {Burrows}, {Lee}, \&
  {Mori}}]{park12}
{Park}, S., {Hughes}, J.~P., {Slane}, P.~O., {et~al.} 2012, \apj, 748, 117

\bibitem[{{Rea} {et~al.}(2016){Rea}, {Borghese}, {Esposito}, {Coti Zelati},
  {Bachetti}, {Israel}, \& {De Luca}}]{rea16}
{Rea}, N., {Borghese}, A., {Esposito}, P., {et~al.} 2016, \apjl, 828, L13

\bibitem[{{Reach} {et~al.}(2006){Reach}, {Rho}, {Tappe}, {Pannuti}, {Brogan},
  {Churchwell}, {Meade}, {Babler}, {Indebetouw}, \& {Whitney}}]{reach06}
{Reach}, W.~T., {Rho}, J., {Tappe}, A., {et~al.} 2006, \aj, 131, 1479

\bibitem[{{Reynoso} {et~al.}(2004){Reynoso}, {Green}, {Johnston}, {Goss},
  {Dubner}, \& {Giacani}}]{reynoso04}
{Reynoso}, E.~M., {Green}, A.~J., {Johnston}, S., {et~al.} 2004, \pasa, 21, 82

\bibitem[{{Safi-Harb} \& {Kumar}(2013)}]{safiharb13}
{Safi-Harb}, S. \& {Kumar}, H.~S. 2013, in IAU Symposium, Vol. 291, Neutron
  Stars and Pulsars: Challenges and Opportunities after 80 years, ed. J.~{van
  Leeuwen}, 480--482

\bibitem[{{S{\'a}nchez-Cruces} {et~al.}(2018){S{\'a}nchez-Cruces}, {Rosado},
  {Fuentes-Carrera}, \& {Ambrocio-Cruz}}]{sanchez-cruces18}
{S{\'a}nchez-Cruces}, M., {Rosado}, M., {Fuentes-Carrera}, I., \&
  {Ambrocio-Cruz}, P. 2018, \mnras, 473, 1705

\bibitem[{{Sasaki} {et~al.}(2013){Sasaki}, {Plucinsky}, {Gaetz}, \&
  {Bocchino}}]{sasaki13}
{Sasaki}, M., {Plucinsky}, P.~P., {Gaetz}, T.~J., \& {Bocchino}, F. 2013, \aap,
  552, A45

\bibitem[{{Sasaki} {et~al.}(2004){Sasaki}, {Plucinsky}, {Gaetz}, {Smith},
  {Edgar}, \& {Slane}}]{sasaki04}
{Sasaki}, M., {Plucinsky}, P.~P., {Gaetz}, T.~J., {et~al.} 2004, \apj, 617, 322

\bibitem[{{Sch{\"o}ller} {et~al.}(2017){Sch{\"o}ller}, {Hubrig}, {Fossati},
  {Carroll}, {Briquet}, {Oskinova}, {J{\"a}rvinen}, {Ilyin}, {Castro}, {Morel},
  {Langer}, {Przybilla}, {Nieva}, {Kholtygin}, {Sana}, {Herrero}, {Barb{\'a}},
  {de Koter}, \& {BOB Collaboration}}]{scholler17}
{Sch{\"o}ller}, M., {Hubrig}, S., {Fossati}, L., {et~al.} 2017, \aap, 599, A66

\bibitem[{{Sedov}(1959)}]{sedov59}
{Sedov}, L.~I. 1959, {Similarity and Dimensional Methods in Mechanics}
  (Similarity and Dimensional Methods in Mechanics, New York: Academic Press,
  1959)

\bibitem[{{Shull} {et~al.}(1985){Shull}, {Dyson}, {Kahn}, \& {West}}]{schull85}
{Shull}, Jr., P., {Dyson}, J.~E., {Kahn}, F.~D., \& {West}, K.~A. 1985, \mnras,
  212, 799

\bibitem[{{Shultz} {et~al.}(2018){Shultz}, {Wade}, {Rivinius}, {Neiner},
  {Alecian}, {Bohlender}, {Monin}, {Sikora}, {MiMeS Collaboration}, \&
  {BinaMIcS Collaboration}}]{shultz18}
{Shultz}, M.~E., {Wade}, G.~A., {Rivinius}, T., {et~al.} 2018, \mnras, 475,
  5144

\bibitem[{{Smartt}(2009)}]{smartt09}
{Smartt}, S.~J. 2009, \araa, 47, 63

\bibitem[{{Soker} \& {Gilkis}(2017)}]{soker17}
{Soker}, N. \& {Gilkis}, A. 2017, \apj, 851, 95

\bibitem[{{Sukhbold} {et~al.}(2016){Sukhbold}, {Ertl}, {Woosley}, {Brown}, \&
  {Janka}}]{sukhbold16}
{Sukhbold}, T., {Ertl}, T., {Woosley}, S.~E., {Brown}, J.~M., \& {Janka}, H.-T.
  2016, \apj, 821, 38

\bibitem[{{Sukhbold} \& {Woosley}(2014)}]{sukhbold14}
{Sukhbold}, T. \& {Woosley}, S.~E. 2014, \apj, 783, 10

\bibitem[{{Taylor}(1950)}]{taylor50}
{Taylor}, G. 1950, Royal Society of London Proceedings Series A, 201, 159

\bibitem[{{Temim} {et~al.}(2019){Temim}, {Slane}, {Sukhbold}, {Koo}, {Raymond},
  \& {Gelfand}}]{temim19}
{Temim}, T., {Slane}, P., {Sukhbold}, T., {et~al.} 2019, arXiv e-prints,
  arXiv:1905.02849

\bibitem[{{Thielemann} {et~al.}(1996){Thielemann}, {Nomoto}, \&
  {Hashimoto}}]{thielemann96}
{Thielemann}, F.-K., {Nomoto}, K., \& {Hashimoto}, M.-A. 1996, \apj, 460, 408

\bibitem[{{Thompson} \& {Duncan}(1993)}]{thompson93}
{Thompson}, C. \& {Duncan}, R.~C. 1993, \apj, 408, 194

\bibitem[{{Thompson} {et~al.}(2004){Thompson}, {Chang}, \&
  {Quataert}}]{thompson04}
{Thompson}, T.~A., {Chang}, P., \& {Quataert}, E. 2004, \apj, 611, 380

\bibitem[{{Tian} \& {Leahy}(2008)}]{tian08}
{Tian}, W.~W. \& {Leahy}, D.~A. 2008, \apj, 677, 292

\bibitem[{{Tong} {et~al.}(2016){Tong}, {Wang}, {Liu}, \& {Xu}}]{tong16}
{Tong}, H., {Wang}, W., {Liu}, X.~W., \& {Xu}, R.~X. 2016, \apj, 833, 265

\bibitem[{{Truelove} \& {McKee}(1999)}]{truelove99}
{Truelove}, J.~K. \& {McKee}, C.~F. 1999, \apjs, 120, 299

\bibitem[{{Vink}(2008)}]{vink08c}
{Vink}, J. 2008, Advances in Space Research, 41, 503

\bibitem[{{Vink} \& {Kuiper}(2006)}]{vink06c}
{Vink}, J. \& {Kuiper}, L. 2006, \mnras, 370, L14

\bibitem[{{Wade} {et~al.}(2014){Wade}, {Grunhut}, {Alecian}, {Neiner},
  {Auri{\`e}re}, {Bohlender}, {David-Uraz}, {Folsom}, {Henrichs}, {Kochukhov},
  {Mathis}, {Owocki}, {Petit}, \& {Petit}}]{wade14}
{Wade}, G.~A., {Grunhut}, J., {Alecian}, E., {et~al.} 2014, in IAU Symposium,
  Vol. 302, Magnetic Fields throughout Stellar Evolution, ed. P.~{Petit},
  M.~{Jardine}, \& H.~C. {Spruit}, 265--269

\bibitem[{{Wilms} {et~al.}(2000){Wilms}, {Allen}, \& {McCray}}]{wilms00}
{Wilms}, J., {Allen}, A., \& {McCray}, R. 2000, \apj, 542, 914

\bibitem[{{Woosley}(2010)}]{woosley10}
{Woosley}, S.~E. 2010, \apjl, 719, L204

\bibitem[{{Woosley} \& {Weaver}(1995)}]{woosley95}
{Woosley}, S.~E. \& {Weaver}, T.~A. 1995, \apjs, 101, 181

\bibitem[{{Xu} \& {Li}(2019)}]{xu19}
{Xu}, K. \& {Li}, X.-D. 2019, \apj, 877, 138

\bibitem[{{Yamane} {et~al.}(2018){Yamane}, {Sano}, {van Loon}, {Filipovi{\'c}},
  {Fujii}, {Tokuda}, {Tsuge}, {Nagaya}, {Yoshiike}, {Grieve}, {Voisin},
  {Rowell}, {Indebetouw}, {Laki{\'c}evi{\'c}}, {Temim}, {Staveley-Smith},
  {Rho}, {Long}, {Park}, {Seok}, {Mizuno}, {Kawamura}, {Onishi}, {Inoue},
  {Inutsuka}, {Tachihara}, \& {Fukui}}]{yamane18}
{Yamane}, Y., {Sano}, H., {van Loon}, J.~T., {et~al.} 2018, \apj, 863, 55

\bibitem[{{Yamauchi} {et~al.}(2008){Yamauchi}, {Ueno}, {Koyama}, \&
  {Bamba}}]{yamauchi08}
{Yamauchi}, S., {Ueno}, M., {Koyama}, K., \& {Bamba}, A. 2008, \pasj, 60, 1143

\bibitem[{{Zhou} \& {Vink}(2018)}]{zhou18a}
{Zhou}, P. \& {Vink}, J. 2018, \aap, 615, A150

\end{thebibliography}

\end{document}